\def\msol{\rm{M}$_{\odot}$}

\def\HII{H\,{\sc ii}}
\def\HI{H\,{\sc i}}
\def\arcsec{$^{\prime}$$^{\prime}$}
\def\arcmin{$^{\prime}$}
\def\deg{$^{\circ}$}
\def\micron{$\mu$m}
\def\etal{\textit{et al.}}

\def\an{AN}

\documentclass{aa}

\usepackage{longtable,lscape,graphicx,float,txfonts,natbib,rotating}

\begin{document}

\title{The RMS Survey: Far-Infrared Photometry of Young Massive Stars.\thanks{The full version of table \ref{T:data_example} is only available in electronic form at the CDS via anonymous ftp to cdsarc.u-strasbg.fr (130.79.125.5) or via http://cdsweb.u-strasbg.fr/cgi-bin/qcat?J/A+A/.}}

\author{J.~C.~Mottram\inst{1,2}~\thanks{E-mail:joe@astro.ex.ac.uk}
  \and
  ~M.~G.~Hoare\inst{1}
  \and
  ~S.~L.~Lumsden\inst{1}
  \and
  ~R.~D.~Oudmaijer\inst{1}
  \and
  ~J.~S.~Urquhart\inst{1,3}
  \and
  ~M.~R.~Meade\inst{4}
  \and
  ~T.~J.~T.~Moore\inst{5}
  \and
  ~J.~J.~Stead\inst{1}
}

\institute{School of Physics and Astronomy, University of Leeds, Leeds, LS2 9JT, UK
  \and
Department of Physics and Astronomy, University of Exeter, Exeter, Devon, EX4 4QL, UK  
  \and
Australia Telescope National Facility, CSIRO, Sydney, NSW 2052, Australia
  \and
Univ. of Wisconsin - Madison, Dept. of Astronomy, 475 N. Charter St., Madison, WI 53716, USA
 \and
Astrophysics Research Institute, Liverpool John Moores University, Twelve Quays House, Egerton Wharf, Birkenhead, CH41 1LD, UK
}

\date{Received XXX 2009 / Accepted XXX 2009}

\abstract
   {The Red MSX Source (RMS) survey is a multi-wavelength campaign of follow-up observations of a colour-selected sample of candidate massive young stellar objects (MYSOs) in the galactic plane. This survey is returning the largest well-selected sample of MYSOs to date, while identifying other dust contaminant sources with similar mid-infrared colours including  a large number of new ultra-compact (UC) \HII{} regions.}
   {To measure the far-infrared (IR) flux, which lies near the peak of the spectral energy distribution (SED) of MYSOs and UC\HII{} regions, so that, together with distance information, the luminosity of these sources can be obtained.}
   {Less than 50~$\%$ of RMS sources are associated with IRAS point sources with detections at 60~\micron{} and 100~\micron{}, though the vast majority are visible in Spitzer MIPSGAL or IRAS Galaxy Atlas (IGA) images. However, standard aperture photometry is not appropriate for these data due to crowding of sources and strong spatially variable far-IR background emission in the galactic plane. A new technique using a 2-dimensional fit to the background in an annulus around each source is therefore used to obtain far-IR photometry for young RMS sources.}
   {Far-IR fluxes are obtained for a total of 1113 RMS candidates identified as young sources. Of these 734 have flux measurements using IGA 60~\micron{} and 100~\micron{} images and 724 using MIPSGAL 70~\micron{} images, with 345 having measurements in both data sets.}
   {}

\keywords{Stars: Formation - Stars: Pre-Main Sequence - \HII{} Regions - Infrared: General - Techniques: Photometric - Surveys}

\authorrunning{J.~C.~Mottram et al.}
\titlerunning{Far-IR photometry of young massive stars}
\maketitle

\section{Introduction}
\label{S:intro}

\begin{figure}
\centering
\includegraphics[width=0.49\textwidth]{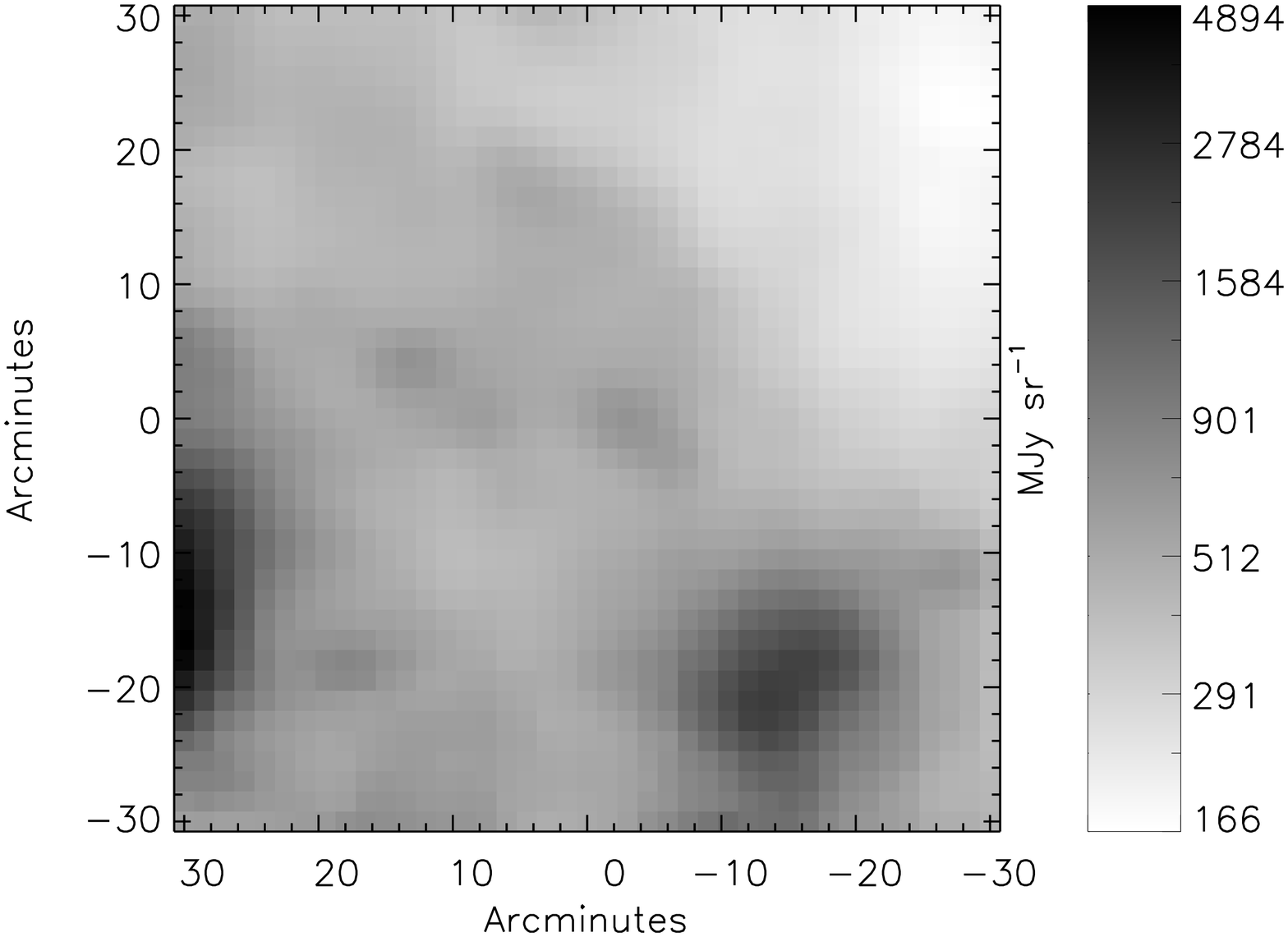}
\includegraphics[width=0.49\textwidth]{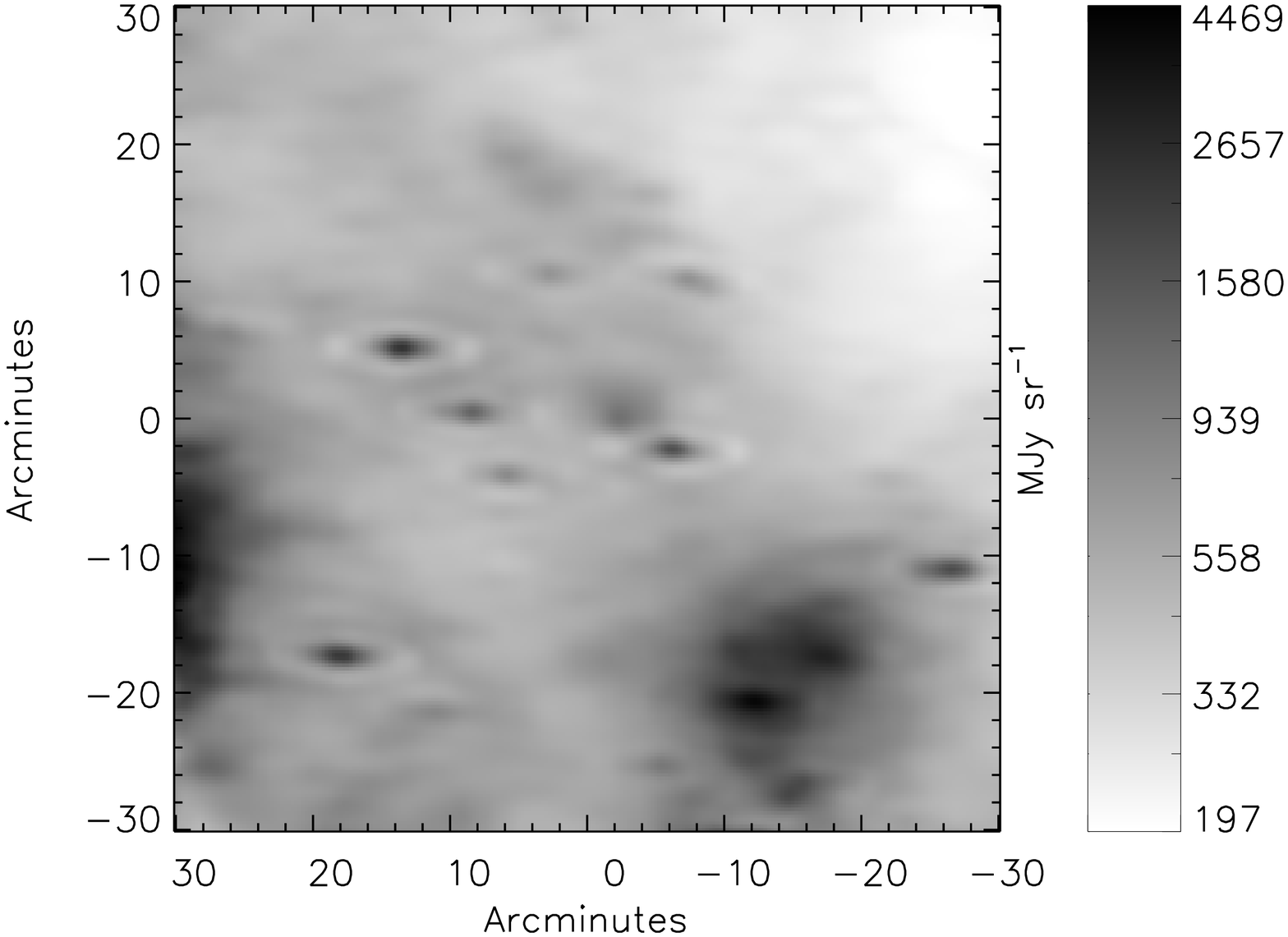}
\includegraphics[width=0.49\textwidth]{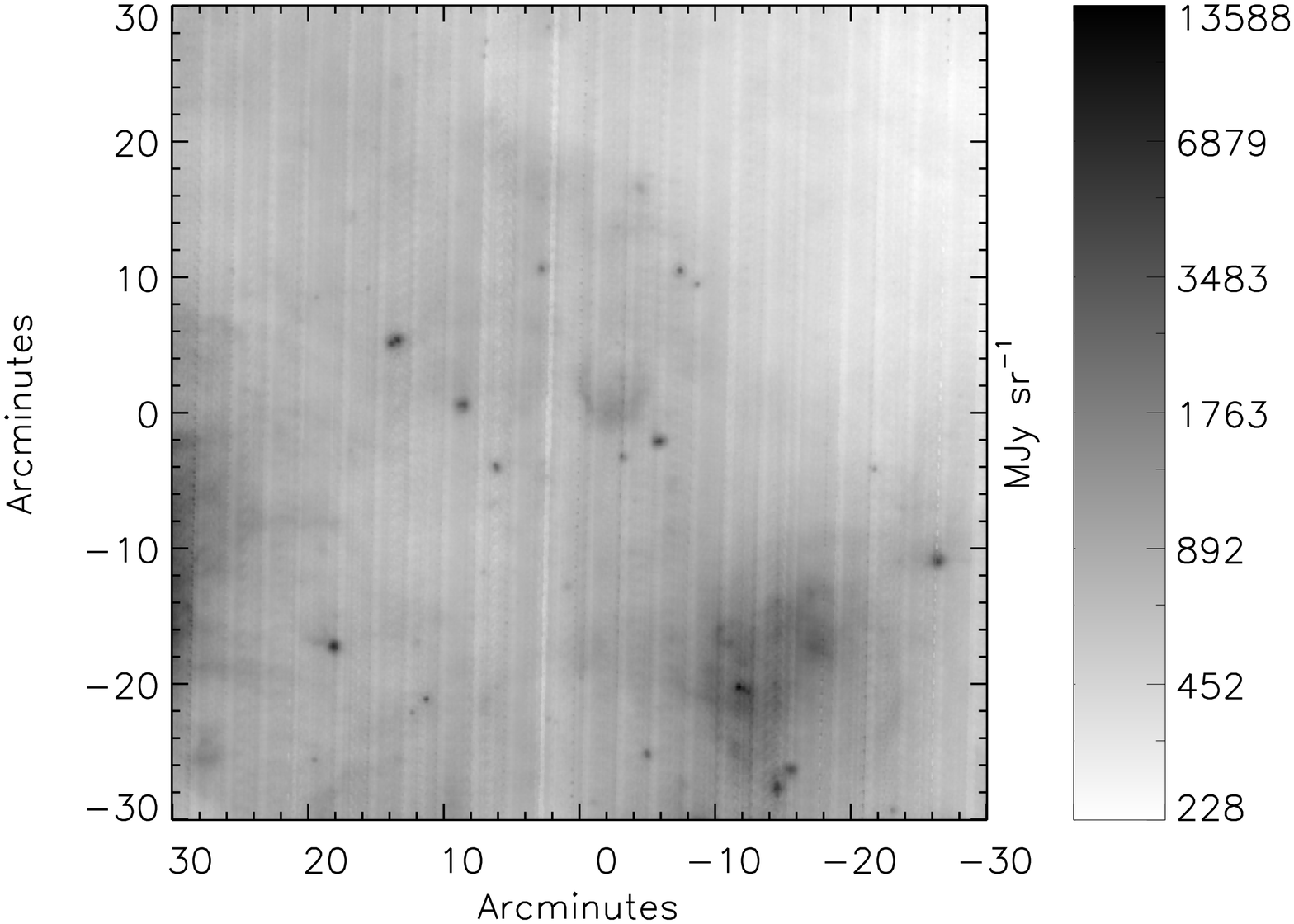}
\caption{A comparison of an IRAS ISSA 60~\micron{} image (top), an IGA 60~\micron{} image after 20 HIRES iterations (centre) and a Spitzer MIPSGAL 70~\micron{} (bottom) image of the same region centred on G015.00+00.00. The relevant features of IGA and MIPSGAL data are discussed in \S\ref{S:data_iga} and \S\ref{S:data_mipsgal}.}
\label{F:data_iga_comparison}
\end{figure}

Massive (M~$\geq$~8~\msol) stars dominate the regions they form in due to their strong ionising radiation and their powerful outflows and winds. However, understanding the early stages of their formation and evolution remains difficult due to their rarity, the embeddedness of the formation process and the short timescale over which it takes place. Of particular interest is the massive young stellar object (MYSO) phase, which is mid-infrared (IR) bright but radio quiet, where core hydrogen burning has probably begun but an \HII{} region has yet to form. As such, MYSOs represent one of the earliest phases of the life of a massive star where it is truly massive while major accretion may still be ongoing, as evidenced by their powerful bipolar molecular outflows \citep[e.g.][]{Beuther2002c,Beuther2002b}.

A major stumbling block to studying MYSOs, and by extension massive star formation in general, is the lack of large well-selected samples of sources from which to differentiate the global properties of all such sources from individual peculiarities and variations. Until recently, the relatively small catalogues of primarily serendipitously discovered MYSOs by \citet{Wynn-Williams1982} and \citet{Henning1984} represented the largest collection of MYSOs but, with only of order 50 sources between them, this is a factor of 10 smaller than expected \citep[][]{Lumsden2002}. This led \citet{Lumsden2002} to identify a sample of $\sim$~2000 candidate MYSOs from the Midcourse Space Experiment (MSX) point source catalogue \citep{Egan1999,Egan2003b} using colour selection criteria.

 The Red MSX Source (RMS) survey is a campaign of follow-up observations designed to identify other dusty objects which contaminate this sample of MYSO candidates (ultra compact (UC) \HII{} regions, low-mass YSOs, Evolved stars, proto-planetary nebulae (PNe) and PNe) and gain information about the sources \citep{Hoare2005,Mottram2006,Urquhart2008b}. These include; $\sim$~1\arcsec{} resolution continuum radio observations to identify regions of ionised emission \citep{Urquhart2007a,Urquhart2009}; ground-based $\sim$~1\arcsec{} resolution mid-IR imaging where Spitzer IRAC (3.6~\micron{}, 4.5~\micron{}, 5.8~\micron{} and 8.0~\micron{}) imaging from the GLIMPSE survey \citep[][]{Benjamin2003,Churchwell2009} is unavailable to identify multiplicity in the $\sim$~18\arcsec{} MSX beam, obtain better astrometry and ensure that the survey does not bias against YSOs near \HII{} regions \citep[e.g.][]{Mottram2007}; $^{13}$CO molecular line observations in order to obtain kinematic distances \citep{Urquhart2007c,Urquhart2008a}; and near-IR spectroscopy \citep[e.g.][]{Clarke2006} to distinguish between YSOs and evolved stars.

In order to obtain the luminosities of the young massive stars identified by the RMS survey, photometry is required over as much of the spectral energy distribution (SED) as possible. In particular, the SEDs of YSOs and UC\HII{} regions peak in the far-IR between $\sim$~60~\micron{} and $\sim$~120~\micron{}. It is therefore important that far-IR photometry is available for those candidates identified as young, i.e. associated with star formation activity. Despite covering the majority of the sky, the IRAS point source catalogue \citep{Beichman1988} does not contain entries for  $\sim$~28~$\%$ of young RMS sources, and a similar number have only upper limit detections at 60~\micron{} and/or 100~\micron{}. In addition, the background emission at far-IR wavelengths is strong and spatially variable within the galactic plane for IRAS data which adds complexity to background subtraction. Simple aperture photometry, which estimates the background emission using the average within an annulus surrounding the source is therefore unlikely to provide accurate measurements. However, fitting the point spread function (PSF) of the detector is also difficult due to the crowded nature of sources in the galactic plane and interaction between the extended background and the PSF.

In this paper, we begin by discussing in \S\ref{S:data} the RMS sample of massive young stars and the data used to obtain far-IR fluxes for them: HIRES reprocessed IRAS Galaxy Atlas \citep[IGA][]{Cao1997} 60~\micron{} and 100~\micron{} images and MIPSGAL \citep{Carey2009} 70~\micron{} images. In \S\ref{S:2dfitting} we outline a new technique for performing aperture photometry using a 2-D fit to the background within an annulus, rather than a simple average. The results of photometric measurements using this technique are presented in \S\ref{S:results}, and discussed in \S\ref{S:discussion}. We then summarise and reach our conclusions in \S\ref{S:conclusions}.

\section{Data}
\label{S:data}

\subsection{The RMS sample of massive young stars}
\label{S:data_rms}

Using the RMS follow-up observations and data from the literature (e.g. 2MASS, GLIMPSE), all of which are available on the RMS database (http://www.ast.leeds.ac.uk/RMS/), each candidate MYSO has been assigned one of several identifications \citep[see][for more details]{Mottram2008}. Young sources can be assigned `YSO', `\HII{} region', `\HII{}~/~YSO' and `Young~/~Old Star'.

Sources assigned the `YSO' identification are radio-quiet mid-IR point sources associated with $^{13}$CO emission which do not look isolated in near-IR images. Where near-IR spectroscopy is available, these sources usually show \HI{} emission, though not at Case-B ratios \citep{Baker1938}, and can also show H$_{2}$, He\,\textsc{i}, Fe\,\textsc{ii} and [Fe\,\textsc{ii}] emission, as well as CO bandhead emission and/or absorption. `\HII{} region's are either extended in $\sim$1\arcsec{} resolution mid-IR imaging and/or radio loud, with Case-B \HI{} line ratios in the near-IR. `\HII{}~/~YSO' sources either have indications that both a YSO and an \HII{} region are present within the MSX beam (18\arcsec{}), or have conflicting evidence as to their identification with insufficient information currently available to determine between the `YSO' and `\HII{} region' designations. Finally sources identified as `Young~/~Old Star's are unlikely to be \HII{} regions or PNe as they are radio quiet and unresolved in the mid-IR, but have insufficient evidence to determine whether the source is a YSO or an evolved star. As more information becomes available identifications are updated where needed, particularly from our near-IR spectroscopy and the luminosities derived using the data presented in this paper, which will be the subject of a forthcoming paper (Mottram \etal{}, in prep.).

\subsection{The IRAS Galaxy Atlas}
\label{S:data_iga}

The IRAS sky survey scanned most regions of the sky more than once, often at different scan angles, thus additional spatial information is available within the data. \citet{Cao1997} therefore undertook HIRES reprocessing of the original IRAS Sky Survey Atlas (ISSA) images \citep{Wheelock1994} for the galactic plane in order to produce the IGA. These data have improved resolution (1.0\arcmin{}~$\times$~1.7\arcmin{} at 60~\micron{}, 1.7\arcmin{}~$\times$~2.2\arcmin{} at 100~\micron{}) with respect to both the original ISSA, as shown in figure~\ref{F:data_iga_comparison}, and the more recent Improved Reprocessing of the IRAS Survey (IRIS) images \citep{MivilleDeschenes2005}, which have resolutions between 3.5\arcmin{} and 5\arcmin{}. Though the IRIS data are better calibrated due to \citet{MivilleDeschenes2005} using COBE/DIRBE data, the improved resolution of the IGA is crucial when considering regions in the galactic plane.

\begin{figure*}
\centering
\includegraphics[width=0.49\textwidth]{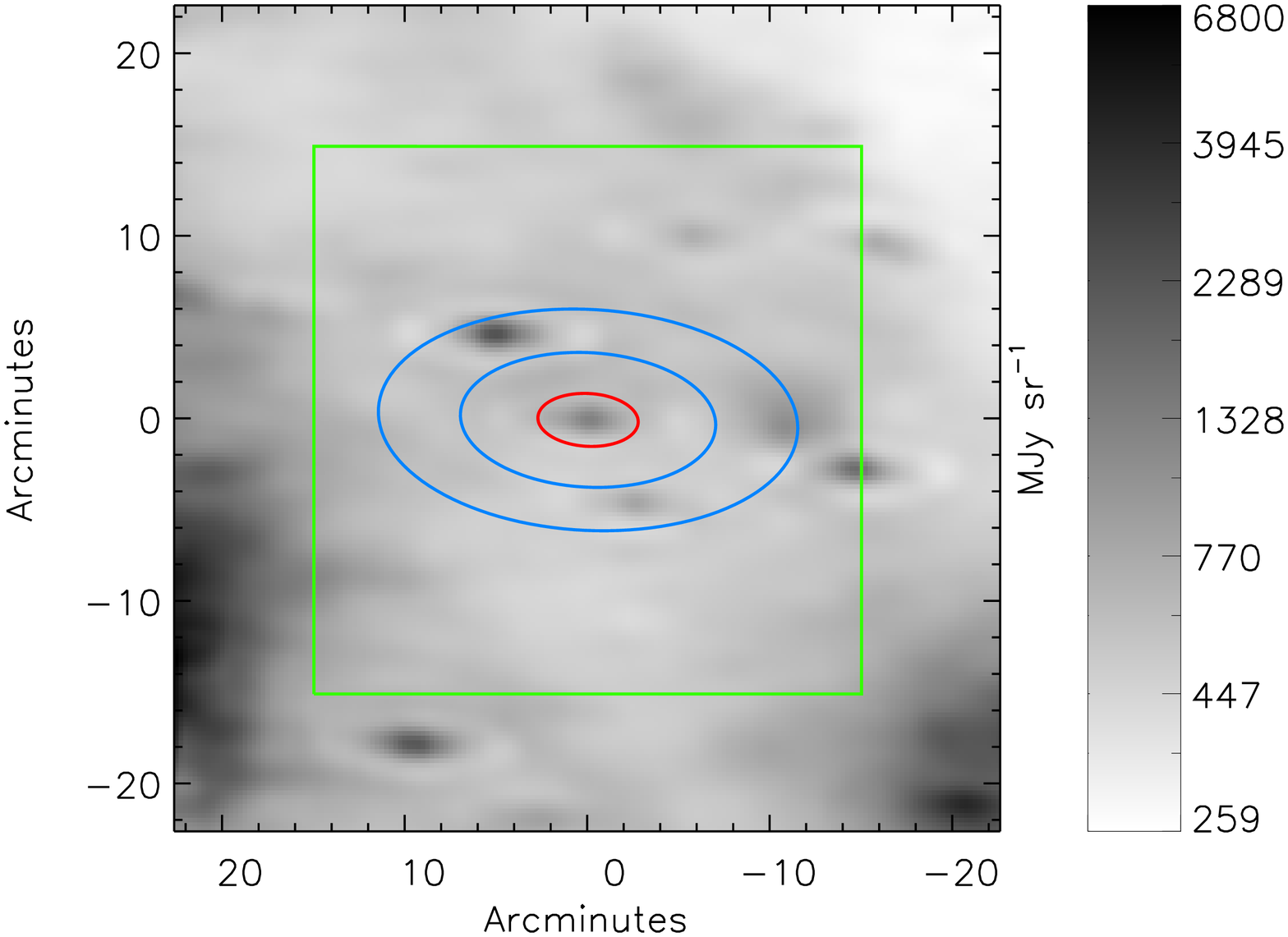}
\includegraphics[width=0.49\textwidth]{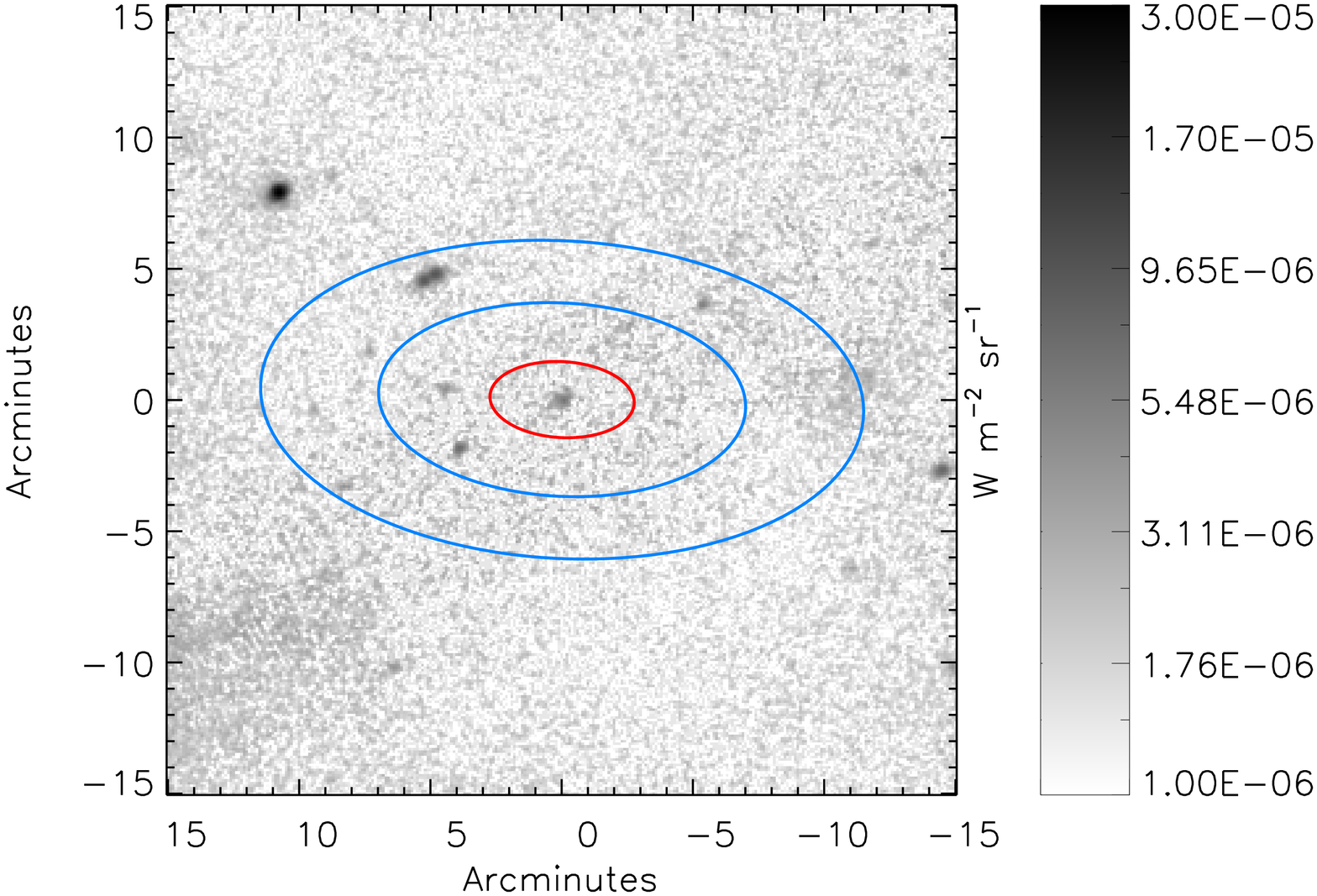}\newline
\includegraphics[width=0.49\textwidth]{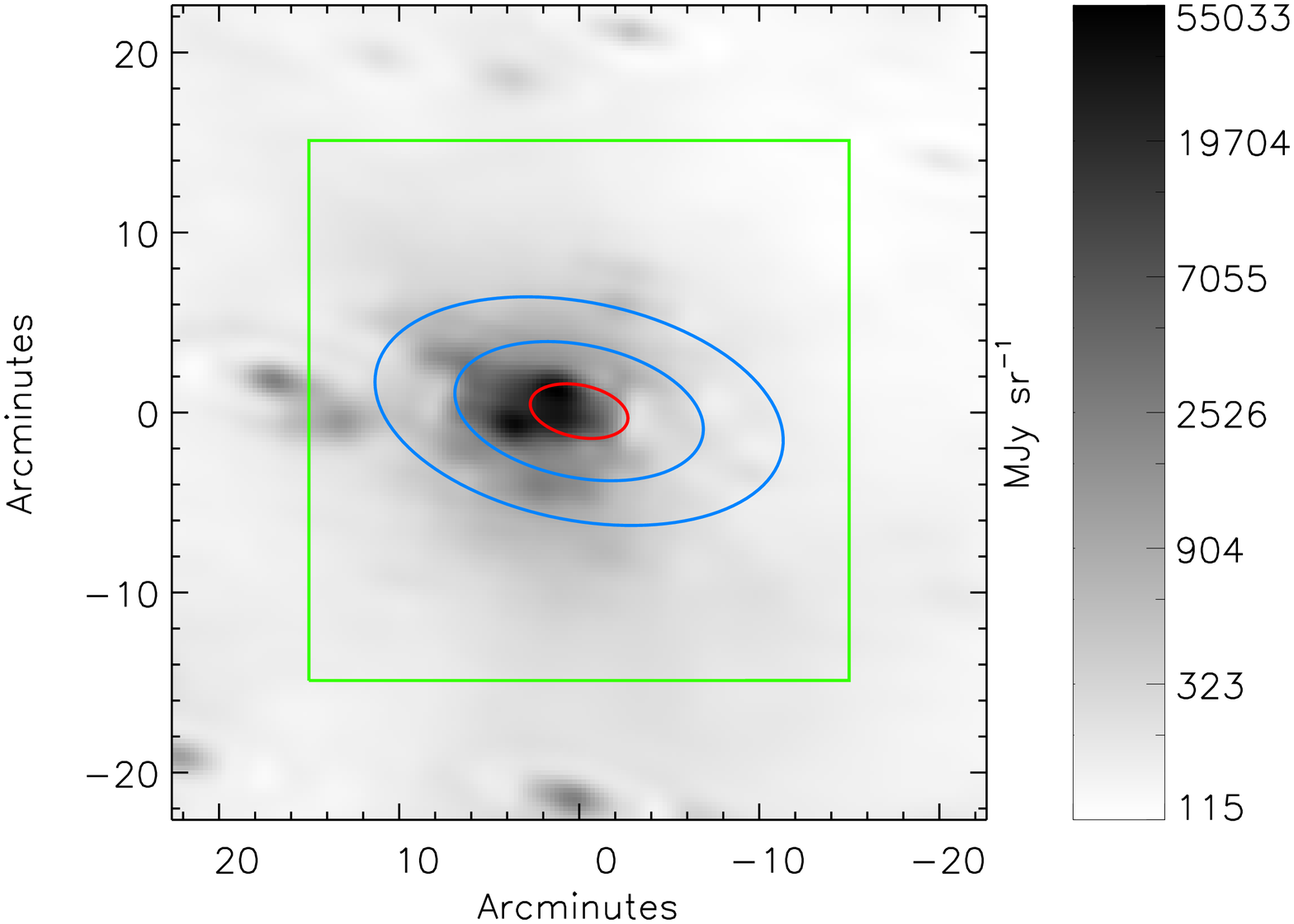}
\includegraphics[width=0.49\textwidth]{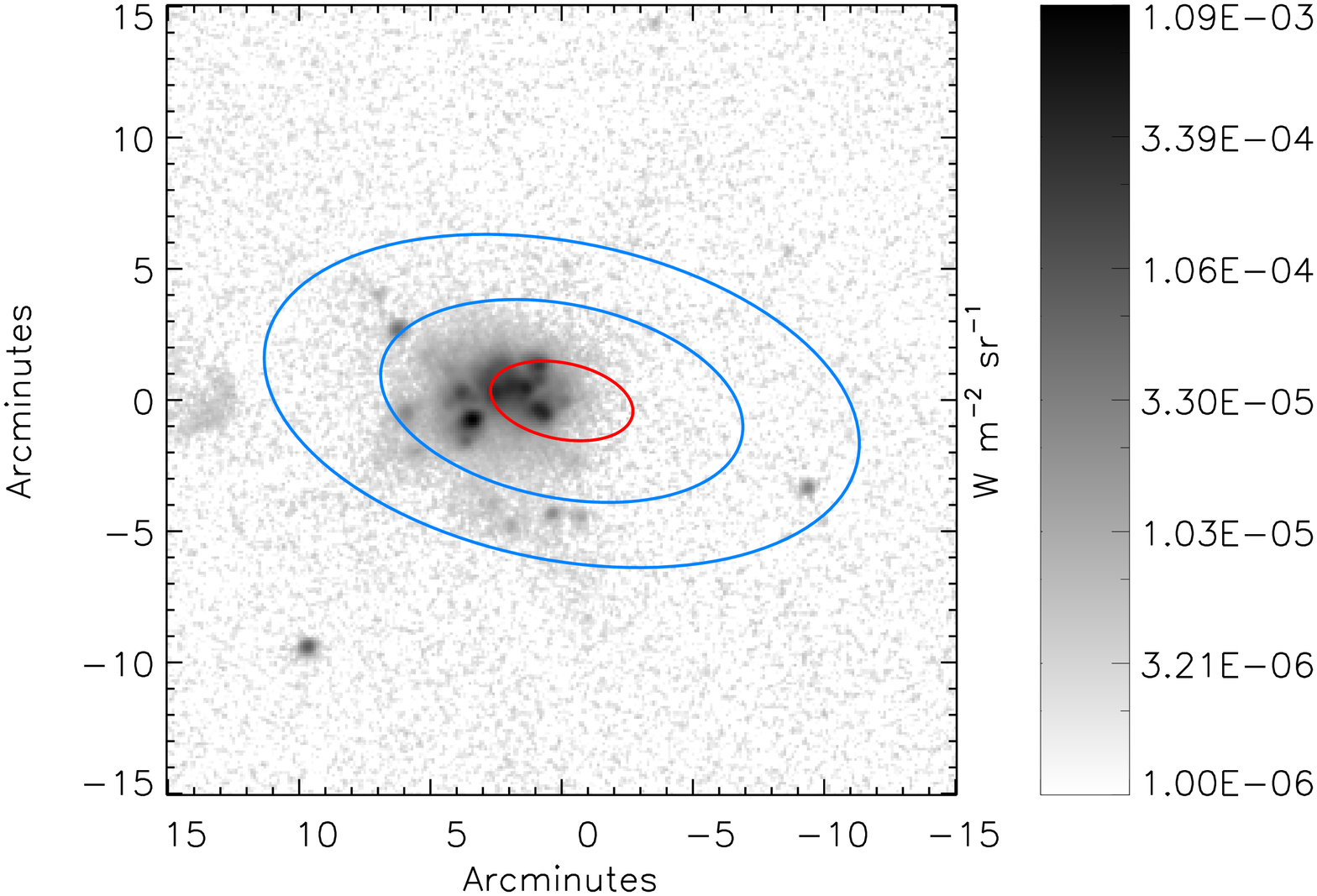}
\caption{Example of a source for which nearby sources do not strongly contaminate the IGA flux (top, G015.0755$-$00.1212) and of a source which is too badly contaminated in the IGA for useful fluxes to be determined (bottom, G043.1497+00.0272). Left: The IGA 60~\micron{} images with 20 HIRES iterations. The standard aperture and sky annulus radii used in two-dimensional background fitting photometry are indicated using ellipses. The box indicates the area covered by the right hand image. Right: MSX E-band (21~\micron{}) image with the IGA 60~\micron{} aperture and sky annulus radii shown as before. For G043.1497+00.0272 the RMS source is just visible on the edge of a large complex of much brighter sources which dominate the flux in the IGA aperture.}
\label{F:data_iga_contexample}
\end{figure*}

Due to the shape of the IRAS detector pixels and the detector layout used, the IRAS sky survey had different resolution parallel and perpendicular to the scanning direction of the satellite. This caused the point source function (PSF) of raw IRAS images to be elliptical, rather than circular $-$ a feature which is retained in the IGA (see figure~\ref{F:data_iga_comparison}), though IRIS images are smoothed to a circular PSF with resolution closer to the larger dimension \citep[i.e. cross-scan, see][]{MivilleDeschenes2005}. The angle between the major axis of the PSF and the $x$-axis of the image also varies across the IGA images due to the changing scan direction of the IRAS satellite across the sky as it orbited the earth and the earth orbited the sun.

As a part of HIRES processing, beam sample map images were produced which have the PSF placed at the same positions on each tile on a smoothed version of the 20 iteration image in order to provide a reasonable estimate of the local background. These were used to measure the local beam angle for each source. In addition, HIRES processing results in a negative ringing artifact around the central maximum of the PSF (see figure~\ref{F:data_iga_comparison}) which must be avoided during aperture photometry, caused by interaction between the PSF and the non-zero background \citep[see][ \S5.2 for a more detailed discussion]{Cao1997}.

Despite the improved resolution of the IGA over the original IRAS data, the resolution is not as good as that of MSX (18.3\arcsec{}) and source contamination may still be an issue (see figure~\ref{F:data_iga_contexample}). Nearby sources in the IRAS point source catalogue (PSC) were identified in order to remove contaminating sources from the sky annulus, and flag sources where the flux measurement is potentially contaminated by other nearby sources. However, the PSC is not necessarily complete within the galactic plane and may not include cases where several sources cluster to give the appearance of a resolved source. Therefore, all sources were subsequently visually inspected and cases where the RMS target was dominated by other nearby sources were flagged as unusable.

\subsection{MIPSGAL}
\label{S:data_mipsgal}

For sources in the inner galactic plane (5\deg{}~$<$~l~$<$63\deg{} and 298\deg{}~$<$~l~$<$355\deg{}) to $\mid$~b~$\mid$~$\lesssim$~1\deg{}, higher resolution data are available from the Spitzer Space Telescope (SST) MIPS Galactic plane (MIPSGAL) survey \citep[][]{Carey2009}, which has a resolution of 18\arcsec{} at 70~\micron{}. Though fully reduced and calibrated mosaic images have not yet been released for this waveband, observational data with basic reduction is available publicly for the whole survey region. This was obtained and mosaiced in galactic coordinates with a pixel scale of 4.8\arcsec{}~pixel$^{-1}$ using the program \textsc{montage} \citep{Berriman2006}. Note that we do not consider the MIPSGAL 24~\micron{} data as the majority of RMS sources are saturated, we already have MSX 21~\micron{} data and this waveband is not crucial to the determination of the total IR flux and therefore luminosity for our sources.

\begin{figure*}
\centering
\includegraphics[width=0.49\textwidth]{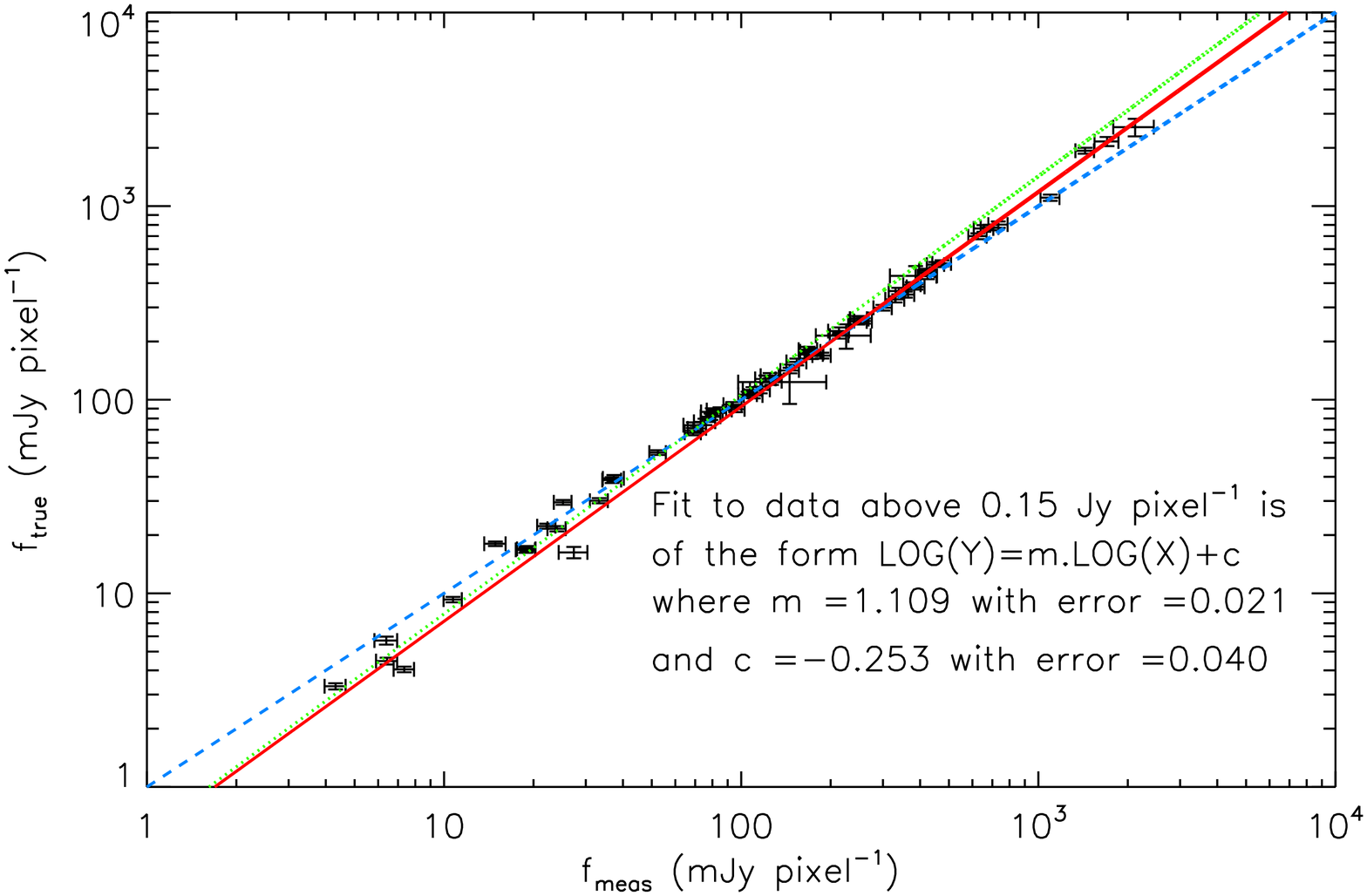}
\includegraphics[width=0.49\textwidth]{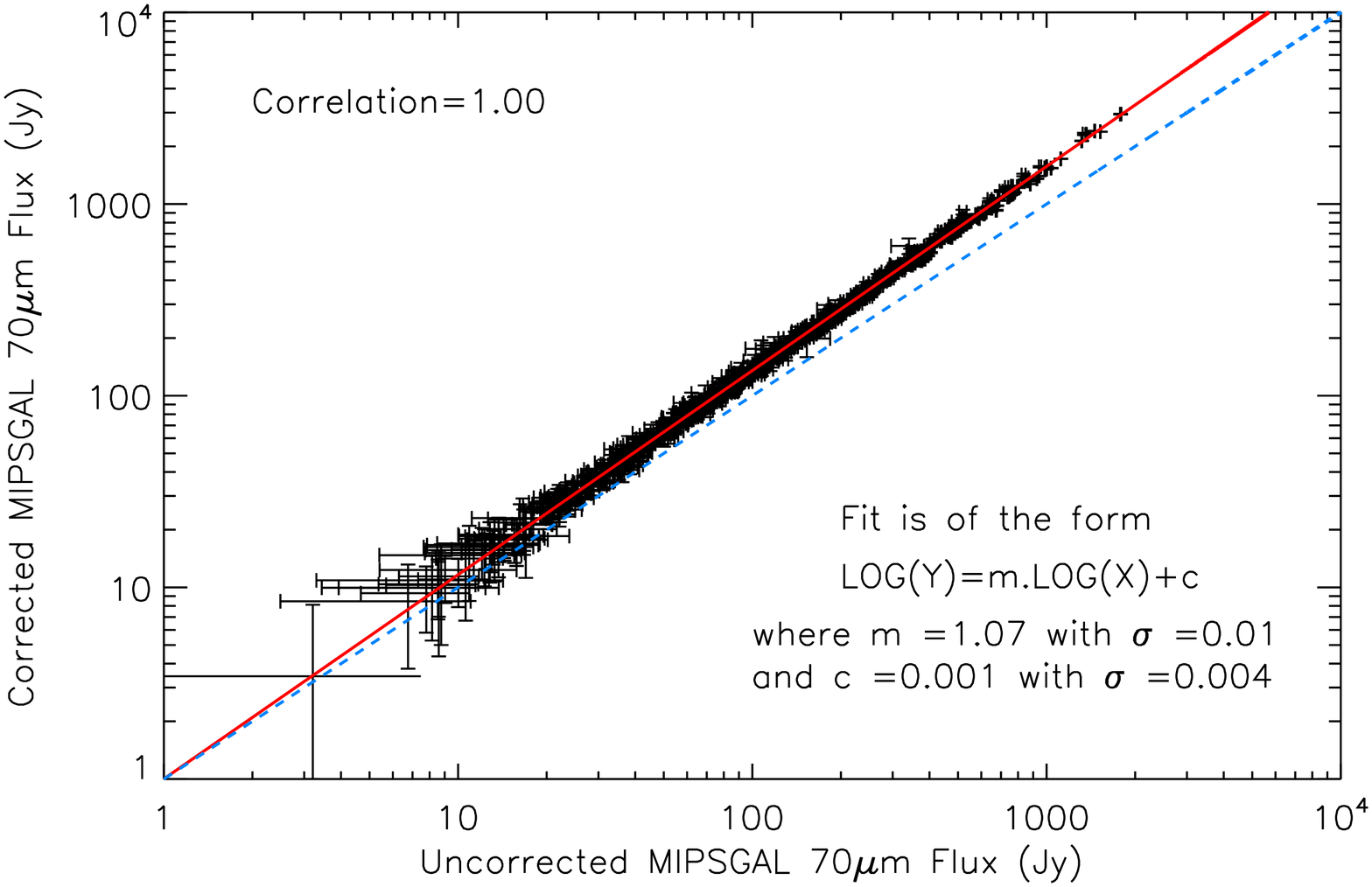}
\caption{Left: The measured and expected pixel surface brightness values derived from the data of \citet{Gordon2007} in order to obtain the MIPS 70~\micron{} pixel surface brightness correction. The solid lines show linear bisector least-squares fit to fluxes $\geq$~1~Jy while the dotted line shows the correction used by \citet{Dale2007} and the dashed line indicates the line of concordance ($y$~=~$x$) and is shown merely to guide the eye. Right: A comparison of corrected and uncorrected MIPSGAL fluxes for all 707 detections. The solid lines show linear bisector least-squares fit to the data which the dashed line again indicates the line of concordance). }
\label{F:data_mipsgal_nonlinear_cor}
\end{figure*}

The MIPS 70~\micron{} detector suffers from a non-linear pixel response for surface brightnesses above $\sim$66~MJy~sr$^{-1}$~\citep{Dale2007} in wide-field mode or 1.5~mJy~pixel$^{-1}$, corresponding to a point source flux of $\sim$1~Jy. Though this is not a particularly large effect in extragalactic studies, the brightness of the galactic plane at these wavelengths results in background emission levels that are often of order 1000~MJy~sr$^{-1}$. We therefore derive a pixel correction using the following method, which is based in that used by \citet[][]{Dale2007}, with the additional inclusion of errors on the data in our linear regression. For each standard star measured using aperture photometry in wide-field observations by \citet[][, their table 4]{Gordon2007}, the measured flux in mJy can be calculated from the measured calibration factor, the predicted flux \citep[derived by][]{Engelbracht2007} and the mean calibration factor obtained from PSF fitting photometry of the same observations. The measured and predicted fluxes are then both converted to surface brightnesses by dividing by 1.5~mJy~pixel$^{-1}$, after which a linear bisector least-squares fit to those sources with fluxes $\geq$~1~Jy (see figure~\ref{F:data_mipsgal_nonlinear_cor}) is used to obtain the pixel correction of the form:

\begin{equation}
F_{true}~=~a(F_{meas})^{b}
\label{E:data_mipsgal_nonlinearity1}
\end{equation}

We find $a$~=~0.558~$\pm$~0.051 and $b$~=~1.11~$\pm$~0.02, which compares well with the results of \citet{Dale2007}. The differences between our correction and that of \citet{Dale2007} come from the use of a linear bisector rather than simply a least squares linear fit. This correction was applied to all target images before photometry was undertaken. Though the MIPSGAL data used in this work has not undergone the post-processing undertaken by \citet{Gordon2007}, this primarily served to counter the slow detector response to the minimal background. RMS sources are generally much brighter and associated with strong background, so this issue is less of a concern. The systematic error introduced by the correction is usually of the order 15$\%$, with the correction in the flux leading to an increase of $\sim$~62~$\%$ at 1000~Jy (see figure \ref{F:data_mipsgal_nonlinear_cor}). Though we are extrapolating the correction to fluxes well above those observed by \citet{Gordon2007}, comparison of corrected MIPS 70\micron{} and IRAS PSC 60\micron{} fluxes (discussed in \S\ref{S:discussion_irasvmips}) returns a slope of 1.00 in log-space, suggesting that the correction is reasonable for all our sources. This is of a simlar order to that expected for such sources from more detailed corrections currently being undertaken by the MIPSGAL team (Paladini, private communication), and the error in this correction dominates any other errors.

As can be seen in figure~\ref{F:data_iga_comparison}, MIPSGAL 70~\micron{} images show striping caused by differences in the background levels between observing scans across the sky. However RMS sources are usually much brighter than the background and striping so the effect is not particularly strong on flux measurements. Therefore no attempts were made to remove it from the images.

\section{Two dimensional background fitting photometry}
\label{S:2dfitting}

In order to perform background subtraction when the background itself is spatially variable (e.g. see figure~\ref{F:data_iga_comparison}), a 2-dimensional fit to the background is the next logical progression from a simple average. As in standard aperture photometry, an annulus around the source of interest is selected, though in this case the surface in the annulus is fitted with a 2-D second order polynomial using a least-squares multiple linear regression method \citep[see][]{Bevington1969}. While a higher dimension fit could be used, it is questionable whether an annulus around a source would uniquely constrain such a fit, or whether the variation in diffuse background truly changes quickly on small spatial scales. Ideally, the background should vary on large scales compared with the size of a point source so that it is fully resolved. 

It is possible, however, that other nearby sources lie partially or wholly within the sky annulus (e.g. see figure~\ref{F:data_iga_contexample}). In such cases the coordinates of these sources can be used to remove pixels up to a given radius around such sources from the area used for the fit, so that they do not affect it. It is not desirable to have too low an area from which to derive the fit, so a cut-off percentage is used below which no more sources are considered for removal.

Having derived the fit to the pixels within the annulus, the fit is then removed from the original image and the total counts within the aperture calculated. The total error in the aperture counts is the combination in quadrature of the noise in the sky within the aperture, the error in the fit to the sky within the annulus and the errors due to the detector (e.g. read noise, gain etc.). With modern detectors it is often difficult to recover this last error explicitly from the released mosaiced images, but it can be considered a limiting minimum error $\sigma_{lim}$, and can therefor be obtained through measurement of regions free from sources of emission. The total error is therefore given by:

\begin{equation}
\sigma_{Tot}=\left(N_{ap}\left(\frac{\Sigma_{an}(c_{o}-c_{f})^{2}}{N_{an}}\right)+\frac{N_{ap}^2}{N_{an}}\left(\frac{\Sigma_{an}(c_{o}-c_{f})^{2}}{N_{an}}\right)+\sigma_{lim}^{2}\right)^{\frac{1}{2}}
\label{E:2dfitting_error}
\end{equation}

\noindent
where $c_{o}$ and $c_{f}$ are the pixel counts of the included pixels within the sky annulus of the original image and 2-D fit respectively while $N_{ap}$ and $N_{an}$ are the number of pixels included in the aperture and annulus.

\begin{figure}
\centering
\includegraphics[width=0.49\textwidth]{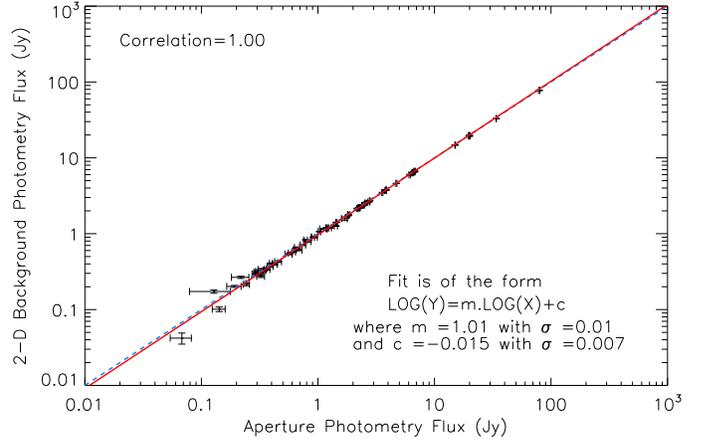}
\caption{Comparison of fluxes measured using aperture photometry and 2-D background fitting photometry for the `clean' subsample of TIMMI2 sources established in \citet{Mottram2007}. The solid line shows a linear bisector least-squares fit to the data including errors while the dashed line indicates the line of equality between the fluxes. The correlation between the data is given in the upper left corner of each plot.}
\label{F:2dfitting_flux}
\vspace{-5mm}
\end{figure}

In order to test this method, the results of aperture photometry and 2-D background fitting photometry were compared using TIMMI2 10.4~\micron{} images from \citet{Mottram2007} of sources in their `clean' subsample (see figure~\ref{F:2dfitting_flux}). The images for these sources have approximately flat uniform backgrounds and isolated unresolved targets, so the results should be the same using either method. The same aperture and inner annulus radii were used as for the standard photometry for TIMMI2 observations (i.e. 12 and 15 pixels respectively) for both methods. However it was found that the 2-D aperture fitting photometry requires a larger sky annulus in order to constrain the fit properly. An outer radius of 30 pixels was therefore used for the aperture fitting photometry, as opposed to the outer radius of 20 pixels used for the standard aperture photometry measurements. The flux was also measured using the 2-D aperture fitting photometry at four locations on each image which were free of source emission and the mean calculated in order to obtain a limiting background flux for each source. The mean of this limiting background flux over all sources was then calculated to obtain the limiting error $\sigma_{lim}$.

The linear bisector least-squares fit to the photometric results for the subsample of TIMMI2 sources (see figure~\ref{F:2dfitting_flux}) has a slope of 1.0 within the errors, and the correlation of aperture photometry to 2-D background fitting photometry fluxes is also 1.0. The two methods are therefore consistent with each other within the calculated errors. In addition to comparisons between the two methods for the same image, comparisons were also performed between the original TIMMI2 fluxes and 2-D background fitting photometry obtained from images which had a diagonal ramp in the background counts and a large bright 2-D gaussian source near the target added to them. These returned very similar results to the first test, so the new method works well even in challenging circumstances.

In order to obtain accurate photometry for IGA data, which has an elliptical beam profile (see figure~\ref{F:data_iga_contexample}),  both standard aperture photometry and aperture fitting photometry routines were also developed to use elliptical instead of circular apertures and annuli.

\section{Results}
\label{S:results}

\subsection{IGA Results}
\label{S:results_iga}

Aperture fitting photometry was undertaken on all 1336 RMS candidate MYSOs identified as young sources, of which 602 were too badly contaminated in IGA data for reliable results to be obtained, leaving 734 sources with acceptable measurements. The aperture and inner and outer sky annulus major axis radii of 11, 28 and 46 pixels were used for the 60~\micron{} data (as shown in figure~\ref{F:data_iga_contexample}) and 13, 30 and 50 pixels for the 100~\micron{} data for all sources, after careful examination of isolated sources. The ratios of the major to minor axis radii used were 1.9 for 60~\micron{} and 1.3 for 100~\micron{}. The limiting flux error $\sigma_{lim}$ was measured using background regions in 13 images for both 60~\micron{} and 100~\micron{} data, resulting in errors of 0.51~Jy and 5.97~Jy respectively. The much larger limiting error for the 100~\micron{} images is probably due to the brightness of diffuse galactic dust emission at this wavelength. 

During their analysis, \citet{Cao1997} found that the IGA after 20 processing iterations overestimates the fluxes of isolated sources with respect to the IRAS PSC. The final flux results therefore include division by the IGA(20)$/$IGA(1) factors measured by \citet{Cao1997} of 1.02 and 1.10 for IGA 60~\micron{} and 100~\micron{} data respectively. This correction results in a systematic error in the fluxes due to the error in measurements of \citet{Cao1997} of 7$\%$ and 14$\%$ in the measured 60~\micron{} and 100~\micron{} fluxes respectively. This systematic error in the fluxes is in addition to the error derived during the 2-D background fitting photometry, which is a random error. Therefore these two errors are kept separate and the systematic error is not included in the results presented in table~\ref{T:data_example}. The correction factors were measured by \citet{Cao1997} using relatively isolated sources with well defined backgrounds, so may well be larger for RMS candidates which are often in complex star formation regions with high levels of background emission.

An example of these results is presented in table~\ref{T:data_example}, while the full version of this table is available at the CDS via anonymous ftp to cdsarc.u-strasbg.fr (130.79.125.5) or via http://cdsweb.u-strasbg.fr/cgi-bin/qcat?J/A+A/. The MSX PSC name for each target is given in column 1 and the IGA measurements and errors are presented in columns 2 and 3 respectively. The coordinates of the nearest IRAS PSC entry within 0.7\arcmin{} (the selection of which will be discussed in \S\ref{S:discussion_irasvmips}) of the MSX coordinates of the target are given for comparison in columns 5 and 6. The offset between the MSX and IRAS PSC coordinates is given in column 7 while the IRAS PSC fluxes with errors are given in columns 8 and 9. The errors in the IGA fluxes were calculated using equation~\ref{E:2dfitting_error}, and are sometimes quite small. However, \citet{Cao1997} estimate that the overall photometric accuracy of the IRAS Galaxy Atlas is $\sim$25$\%$, so the dominant error in these results is probably from the image data rather than the measurements. 

\begin{table*}
\centering
\caption{Example of measured IGA 60~\micron{}, IGA 100~\micron{} and MIPSGAL 70~\micron{} photometry, see text for details. A full version of this table is available online at the CDS via anonymous ftp to cdsarc.u$-$strasbg.fr (130.79.125.5) or via http://cdsweb.u$-$strasbg.fr/cgi$-$bin/qcat?J/A+A/.}
\begin{tabular}{@{}c@{~~}c@{~~~}c@{~~~}c@{~~~}c@{~}c@{~}c@{~~~}c@{~~~}c@{}}
\hline
Object & IGA 60\micron{} & IGA 100\micron{} & MIPSGAL 70\micron{} & \multicolumn{3}{c}{Nearest IRAS PSC Entry} & IRAS 60\micron{} & IRAS 100\micron{} \\
 & Flux (Jy) & Flux (Jy) & Flux (Jy) & RA (J2000) & Dec (J2000) & Offset (\arcmin{}) & Flux (Jy) & Flux (Jy) \\
\hline
G011.0732$-$00.2295&134.48$\pm$2.14&132.32$\pm$12.52&59.35$\pm$3.20&18:10:48.7&$-$19:27:42&0.210&101.00$\pm$12$\%$&294.00$\pm$18$\%$\\
G011.1109$-$00.4001&1145.32$\pm$1.86&2323.30$\pm$10.73&888.59$\pm$3.56&18:11:33.2&$-$19:30:39&0.219&853.00$\pm$16$\%$&2090.00$\pm$17$\%$\\
G011.1723$-$00.0656&181.62$\pm$2.85&299.14$\pm$10.16&114.20$\pm$3.16&18:10:25.2&$-$19:17:51&0.078&73.30$\pm$13$\%$&$<$330.00\\
G011.9454$-$00.0373&1423.15$\pm$3.95&1884.77$\pm$14.14&1119.95$\pm$3.27&18:11:53.0&$-$18:36:20&0.116&1180.00$\pm$17$\%$&1710.00$\pm$15$\%$\\
G012.5932$-$00.5708&94.24$\pm$1.75&229.58$\pm$9.25&17.77$\pm$4.01&18:15:10.0&$-$18:17:30&0.105&$<$57.90&$<$624.00\\
G012.9090$-$00.2607&3212.46$\pm$15.41&8116.61$\pm$27.53&2136.63$\pm$5.53&18:14:39.0&$-$17:52:03&0.183&2230.00$\pm$23$\%$&6310.00$\pm$22$\%$\\
G013.2097$-$00.1436&1214.14$\pm$4.21&2999.06$\pm$14.72&1118.61$\pm$3.59&18:14:47.8&$-$17:32:48&0.511&877.00$\pm$22$\%$&1850.00$\pm$27$\%$\\
G013.6562$-$00.5997&403.97$\pm$2.27&989.68$\pm$9.72&403.22$\pm$3.12&18:17:24.4&$-$17:22:13&0.027&422.00$\pm$21$\%$&1140.00$\pm$17$\%$\\
G013.8885$-$00.4760&661.08$\pm$4.96&861.81$\pm$26.79&169.60$\pm$3.57&18:17:24.7&$-$17:06:26&0.023&262.00$\pm$16$\%$&407.00$\pm$36$\%$\\
G014.0329$-$00.5155&265.45$\pm$16.31&1195.19$\pm$8.10&70.79$\pm$3.64&18:17:50.7&$-$16:59:55&0.040&$<$145.00&923.00$\pm$25$\%$\\
G014.2071$-$00.1105&346.53$\pm$4.85&566.17$\pm$14.71&201.11$\pm$3.48&$\dagger$&$-$&$-$&$-$&$-$\\
G014.2166$-$00.6344&70.15$\pm$4.50&398.38$\pm$13.86&57.22$\pm$3.21&18:18:38.5&$-$16:53:33&0.088&34.40$\pm$18$\%$&$<$2820.00\\
G014.6703$-$00.4795&138.96$\pm$1.63&594.71$\pm$35.98&40.19$\pm$3.47&$\dagger$&$-$&$-$&$-$&$-$\\
G015.0755$-$00.1212&249.51$\pm$4.06&575.18$\pm$14.22&211.48$\pm$3.29&18:18:27.6&$-$15:53:36&0.086&181.00$\pm$20$\%$&$<$477.00\\
G015.0939+00.1913&127.00$\pm$2.36&480.57$\pm$8.71&75.13$\pm$3.14&$\dagger$&$-$&$-$&$-$&$-$\\
G015.4805$-$00.3900&175.60$\pm$1.20&712.90$\pm$6.77&$<$38.80&$\dagger$&$-$&$-$&$-$&$-$\\
G016.1438+00.0074&44.23$\pm$4.26&161.02$\pm$14.41&85.67$\pm$3.17&18:20:03.7&$-$14:53:31&0.260&98.20$\pm$9$\%$&233.00$\pm$17$\%$\\
\hline
\multicolumn{9}{l}{$^\dagger$ No IRAS PSC entry within 0.7\arcmin{} of MSX target coordinates.} \\
\end{tabular}
\label{T:data_example}
\end{table*}

\begin{figure*}
\centering
\includegraphics[width=0.49\textwidth]{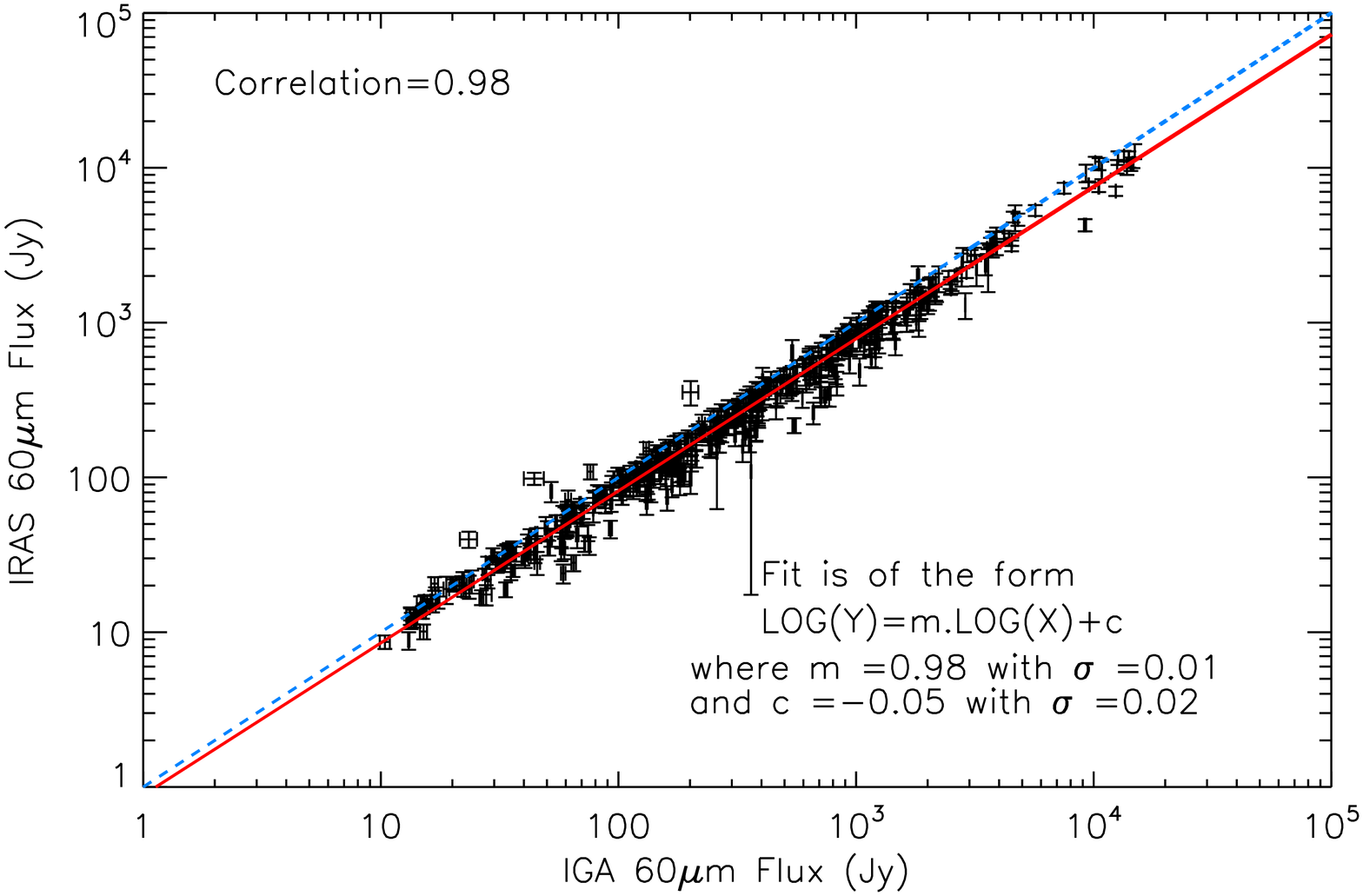}
\includegraphics[width=0.49\textwidth]{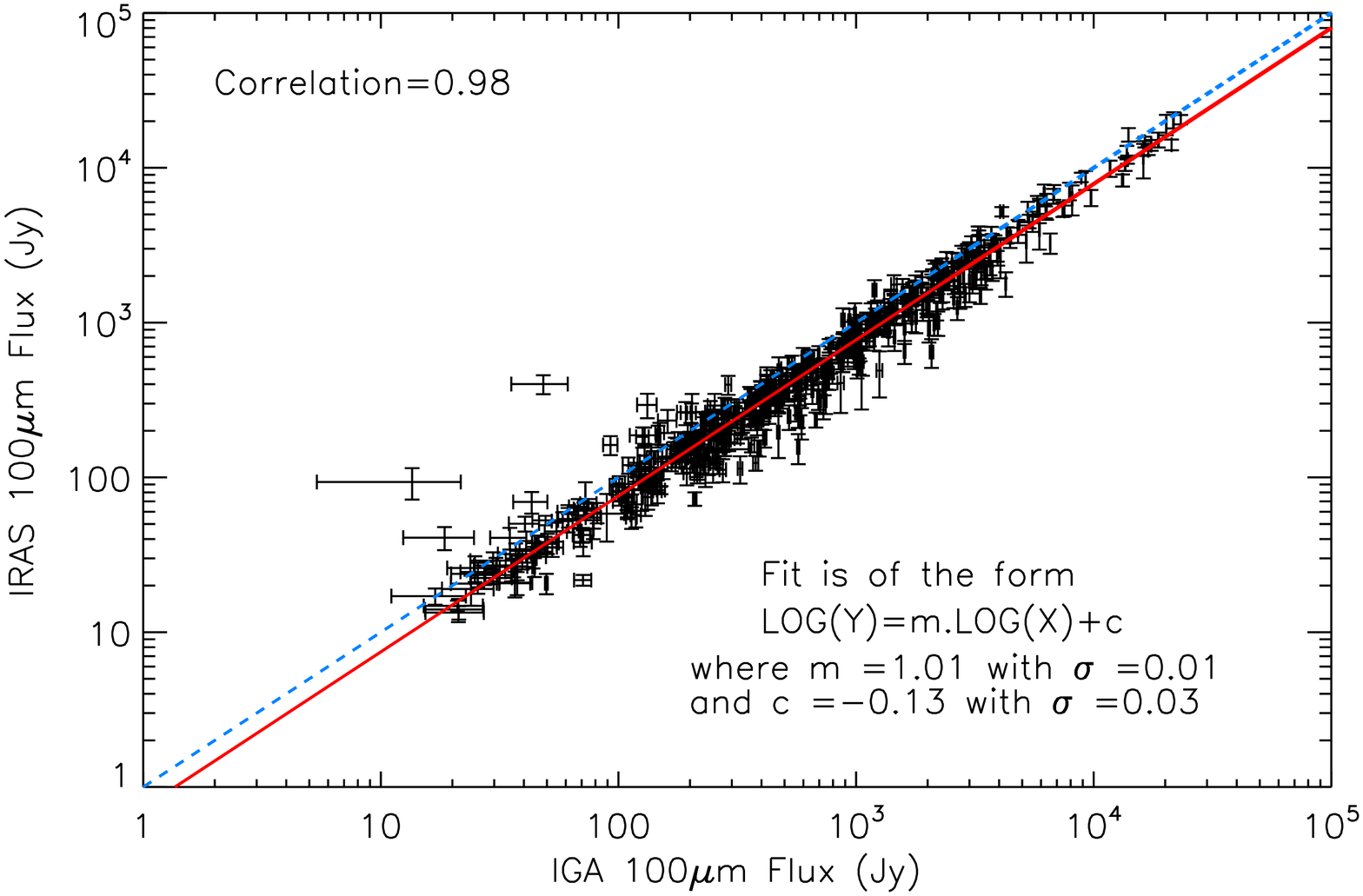}
\caption{A comparison of measured IGA 60~\micron{} fluxes with IRAS 60~\micron{} (left) and IGA 100~\micron{} fluxes with IRAS 100~\micron{} (right) PSC fluxes for 510 sources with PSC detections. The solid line shows the linear-bisector least-squares fit to the data while the dashed line indicates the line of concordance.}
\label{F:iga_results_photiras}
\end{figure*}

Of the 734 targets with IGA fluxes, 682 have an IRAS PSC source within 0.7\arcmin{} but 172 of these PSC sources are upper limit detections at either 60~\micron{}, 100~\micron{} or both. Therefore, IGA fluxes have been obtained for 224 sources which have no far-IR detection in the IRAS PSC at these wavelengths. Figure~\ref{F:iga_results_photiras} shows the comparison between IGA and IRAS PSC fluxes for the 510 sources which have IGA fluxes and true IRAS PSC detections at 60~\micron{} and 100~\micron{}. The IGA fluxes are, on average, slightly larger than the IRAS PSC fluxes by a factor of $\sim$1.12~$\pm$~0.05 at 60~\micron{} and $\sim$1.35~$\pm$~0.09 at 100~\micron{}, however the fluxes in the IRAS PSC did not include a correction for hysteresis at 60~\micron{}, which when included tends to increase these fluxes and would account for the increase seen \citep[][give IGA(1)/PSC~=~1.12]{Cao1997}. The increase in the 100~\micron{} data may be due to flux increases during HIRES processing as our sources are often in regions of considerable background emission which can cause interactions between this and the point source. However, the difference may also be due to more accurate background subtraction.

\subsection{MIPSGAL Results}
\label{S:results_mipsgal}

\begin{figure*}
\centering
\includegraphics[width=0.49\textwidth]{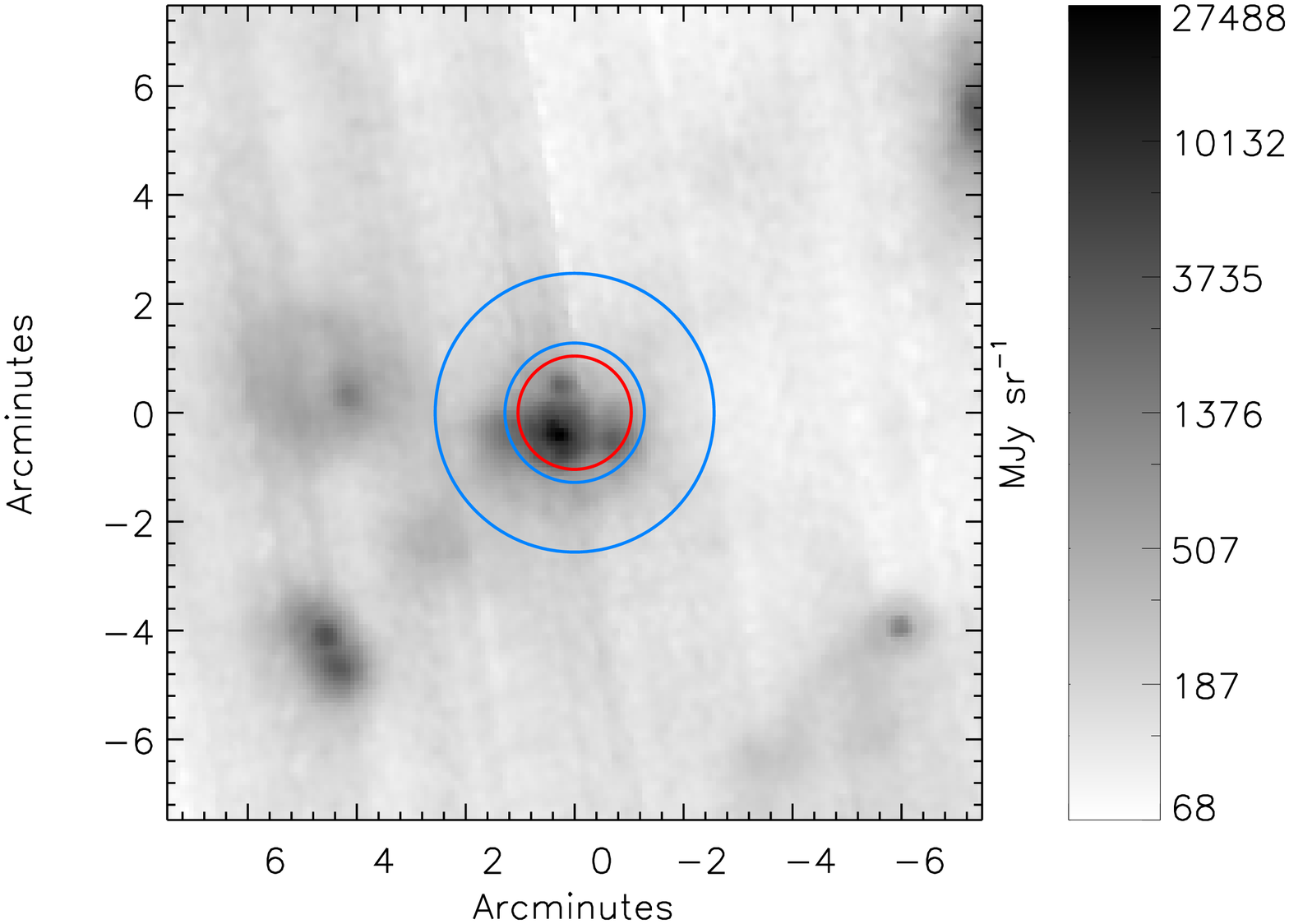}
\includegraphics[width=0.49\textwidth]{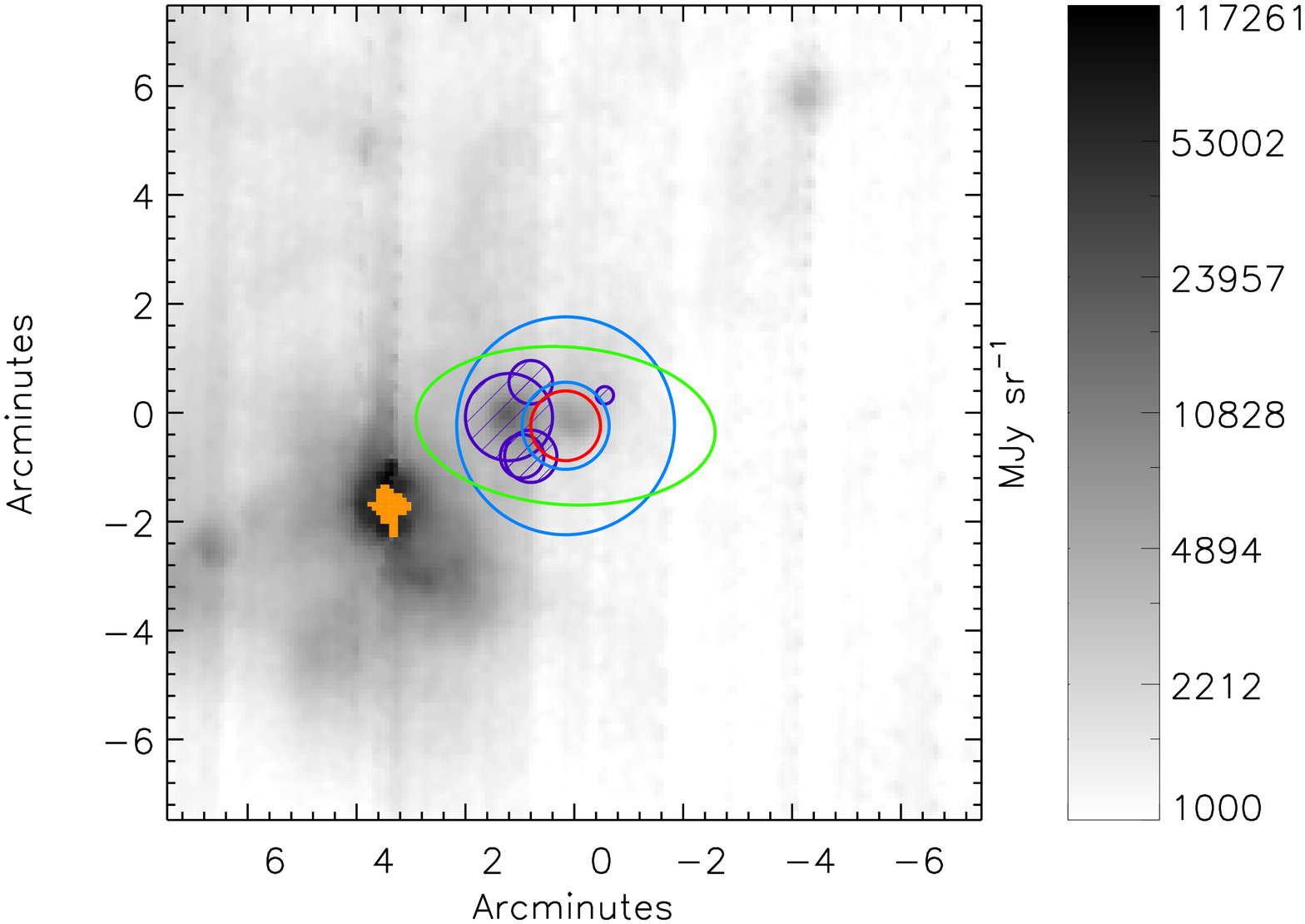}\newline
\caption{Examples of MIPSGAL 70~\micron{} images for RMS targets. The apertures and sky annuli used in photometry are shown with open circles while excluded regions are indicated by line-filled circles. Unphysical pixels are shaded orange in the online version of this paper. Right: a source (G059.7856+00.0734) where nearby bright sources would confuse any flux measurement. Left: a source (G010.6207$-$00.3199) with brighter nearby sources which would be confused in IGA observations (60~\micron{} aperture shown by an ellipse).}
\label{F:results_mipsgal_example}
\end{figure*}

2-D background fitting photometry was attempted for 843 Red MSX Source survey candidate MYSOs identified as young sources which lie within the MIPSGAL survey region. However, 85 were either heavily saturated or have a much brighter source nearby so that no usable information could be obtained from the image (e.g. see the left hand plot of figure~\ref{F:results_mipsgal_example}) and 34 are visible but saturated. Of the 724 sources where flux measurements were possible, 693 are detected and measurable in MIPSGAL images, though nearby sources had to be masked from the sky annulus for some of these sources (e.g. see the right hand plot of figure~\ref{F:results_mipsgal_example}). In addition, 12 sources are not detected at 70~\micron{} and therefore probably not MYSOs or \HII{} regions and 19 sources are non-detections due to there being bright sources or complexes nearby, and so have large upper limits on the flux associated with them. For an isolated unresolved source, aperture and sky radii of 13, 16 and 32 pixels were used, while these radii were set by inspection of the images for more complex sources.

The point source function (PSF) of Spitzer MIPS observations is circular but slightly non-gaussian and includes a low-level extended contribution. An aperture correction was therefore calculated for the MIPSGAL images by performing photometry on the image of an unresolved source with a relatively flat background using various aperture radii and inner and outer sky annulus radii, then dividing the results of photometry with a particularly large aperture which measures as close to the total source flux as possible. This ratio then provides an aperture calibration which is multiplied by the measured flux in order to arrive at the total flux for a given source. Calculation of the aperture correction was undertaken after pixel non-linearity correction of the images (see \S\ref{S:data_mipsgal}). A sample of correction values are shown in table~\ref{T:results_mipsgal_apcorr}, and compare well to those obtained by \citet{Gordon2007} using model PSFs for blackbodies of 10,000~K, 60~K and 10~K, particularly those for the 10~K blackbody. The errors in the calculated corrections are generally below 10$\%$, and are primarily due to the limiting error associated with the measurements used to derive the correction factors. The limiting error $\sigma_{lim}$ was found to be 2.68~Jy, measured as discussed previously, using 17 images with relatively flat backgrounds.

\begin{table}
\centering
\begin{tabular}{@{~}cccc@{~}}
\hline
\centering
Aperture & \multicolumn{2}{c}{Sky Radius} & Aperture \\
Radius (\arcsec{}) & Inner (\arcsec{}) & Outer (\arcsec{}) & Correction \\
\hline
19.2~(4 pix) & 28.8~(6 pix) & 57.6~(12 pix) & 2.50~$\pm$~0.42\\
33.6~(7 pix) & 38.5~(8 pix) & 67.2~(14 pix) & 1.48~$\pm$~0.17\\
38.5~(8 pix) & 48.0~(10 pix) & 96.0~(20 pix) & 1.22~$\pm$~0.12\\
62.4~(13 pix) & 76.8~(16 pix) & 153.6~(32 pix) & 1.14~$\pm$~0.11\\
\hline
\end{tabular}
\caption{Sample MIPSGAL 70~\micron{} aperture correction values calculated using aperture fitting photometry for an isolated unresolved source.}
\label{T:results_mipsgal_apcorr}
\end{table} 

For non-detections, four measurements of the background flux were taken using the standard aperture and annulus radii and the mean of the absolute values of the measurements calculated. A 3~$\sigma$ upper limit on the flux for the source is then given by three times this mean. For those sources which are non-detections due to bright nearby emission, the flux was calculated in the same way as for a standard source, and this flux used as an upper limit to the true MIPSGAL 70~\micron{} flux for the source. Though the saturation limits of MIPSGAL 70~\micron{} images have yet to be fully characterised, it is certainly a problem for bright RMS sources or sources in or near bright regions of the galactic plane.

Example MIPSGAL 70~\micron{} 2-D background fitting photometry results are shown in column 4 of table~\ref{T:data_example}, while the full version of this table is available at the CDS via anonymous ftp to cdsarc.u-strasbg.fr (130.79.125.5) or via http://cdsweb.u-strasbg.fr/cgi-bin/qcat?J/A+A/. All sources which either have a flux measurement or are non-detections for whatever reason are included in the full table.

While the resolution of SST MIPS 70~\micron{} observations is much better than previous surveys (18\arcsec{}), allowing direct comparison with the MSX survey, higher resolution mid-IR imaging of RMS sources \citep[e.g. see][]{Mottram2007} have shown that source confusion may still be an issue. Every effort has been made to ensure that only the source of interest is included within these flux measurements, but it is possible that some unresolved sources may in fact contain more than one emitting object.

\section{Discussion}
\label{S:discussion}

\subsection{Comparison of IRAS and MIPSGAL photometry}
\label{S:discussion_irasvmips}

\begin{figure}
\centering
\includegraphics[width=0.49\textwidth]{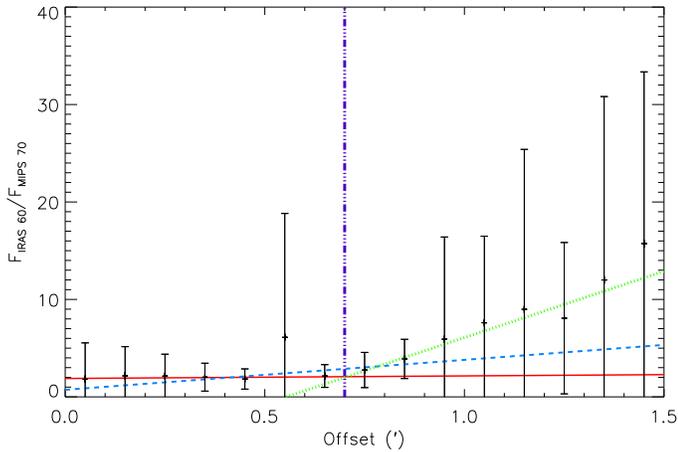}
\caption{The mean ratio of IRAS PSC 60~\micron{} flux to measured MIPSGAL 70~\micron{} flux for the nearest IRAS PSC entry to the target MSX coordinates as a function of offset. The solid line shows the linear bisector least-squares fit to the data below 0.7\arcmin{}, the dotted line shows the fit to the data between 0.7\arcmin{} and 1.5\arcmin{} and the dashed line shows the fit between 0\arcmin{} and 1.5\arcmin{}. The vertical dot-dashed line indicates 0.7\arcmin{}, above which offset the correlation between RMS targets and the IRAS PSC becomes weaker.}
\label{F:discussion_irasvmips_mipsiras}
\end{figure}

In order to better evaluate the MIPSGAL results, a search for IRAS PSC entries associated with RMS sources was conducted. For each source with a MIPSGAL 70~\micron{} detection, the nearest IRAS PSC entry with detections at both 60\micron{} and 100~\micron{} to the MSX coordinates of the RMS source was identified. Figure~\ref{F:discussion_irasvmips_mipsiras} shows the mean IRAS 60~\micron{} to MIPSGAL 70~\micron{} flux ratio as a function of offset between the IRAS and MSX coordinates. From this figure, as well as visual inspection of MSX and MIPSGAL images, the radius out to which the nearest IRAS PSC candidate is considered a probable counterpart to the RMS candidate was set at 0.7\arcmin{}. This is more than twice the resolution of MSX (18.3\arcsec{}) and almost three times the quoted IRAS PSC cross-scan positional error (15.6\arcsec{}). This limit was also used to identify probable IRAS counterparts for sources with IGA observations (see \S\ref{S:results_iga}).

\begin{figure*}
\centering
\includegraphics[width=0.49\textwidth]{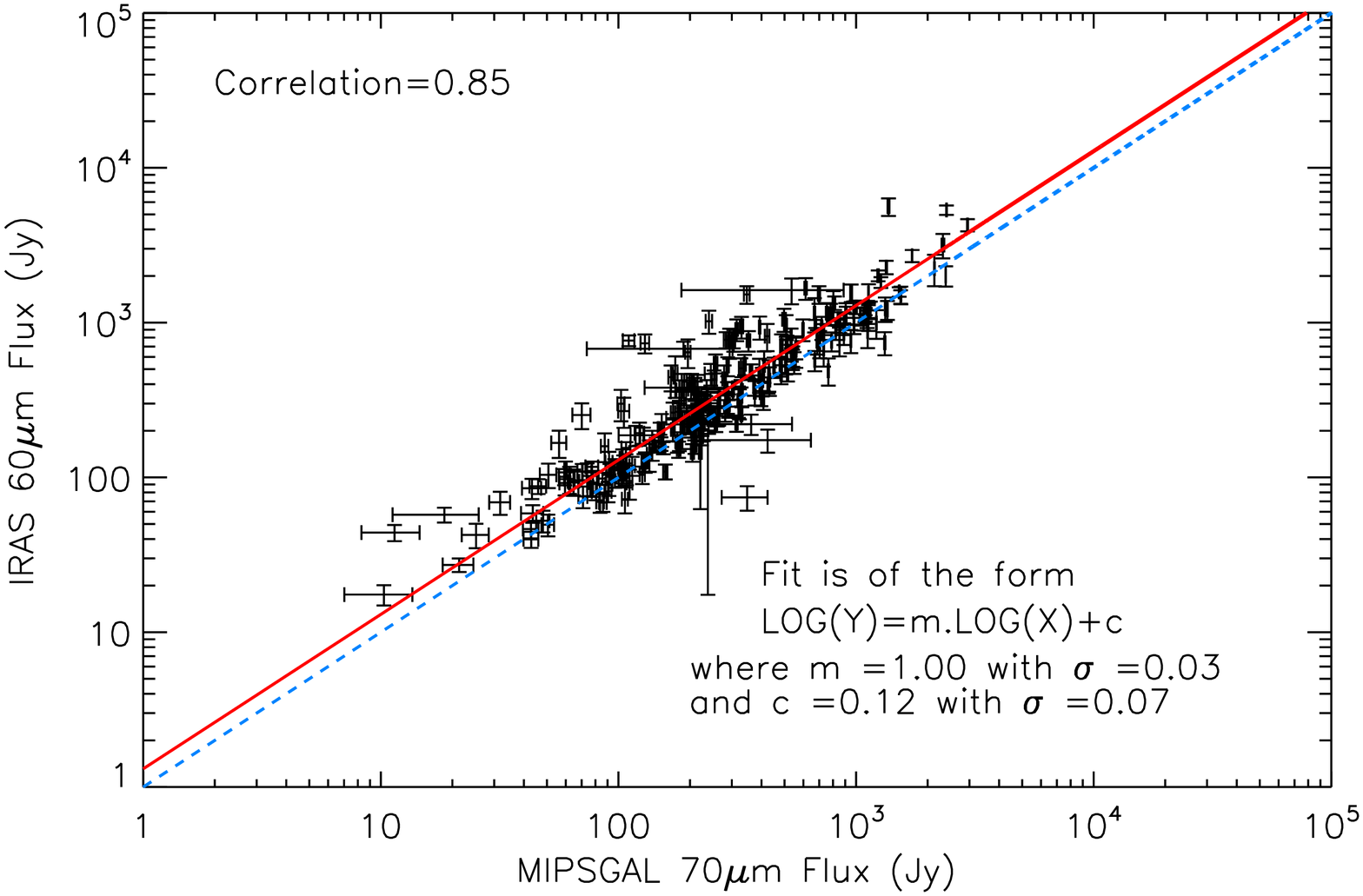}
\includegraphics[width=0.49\textwidth]{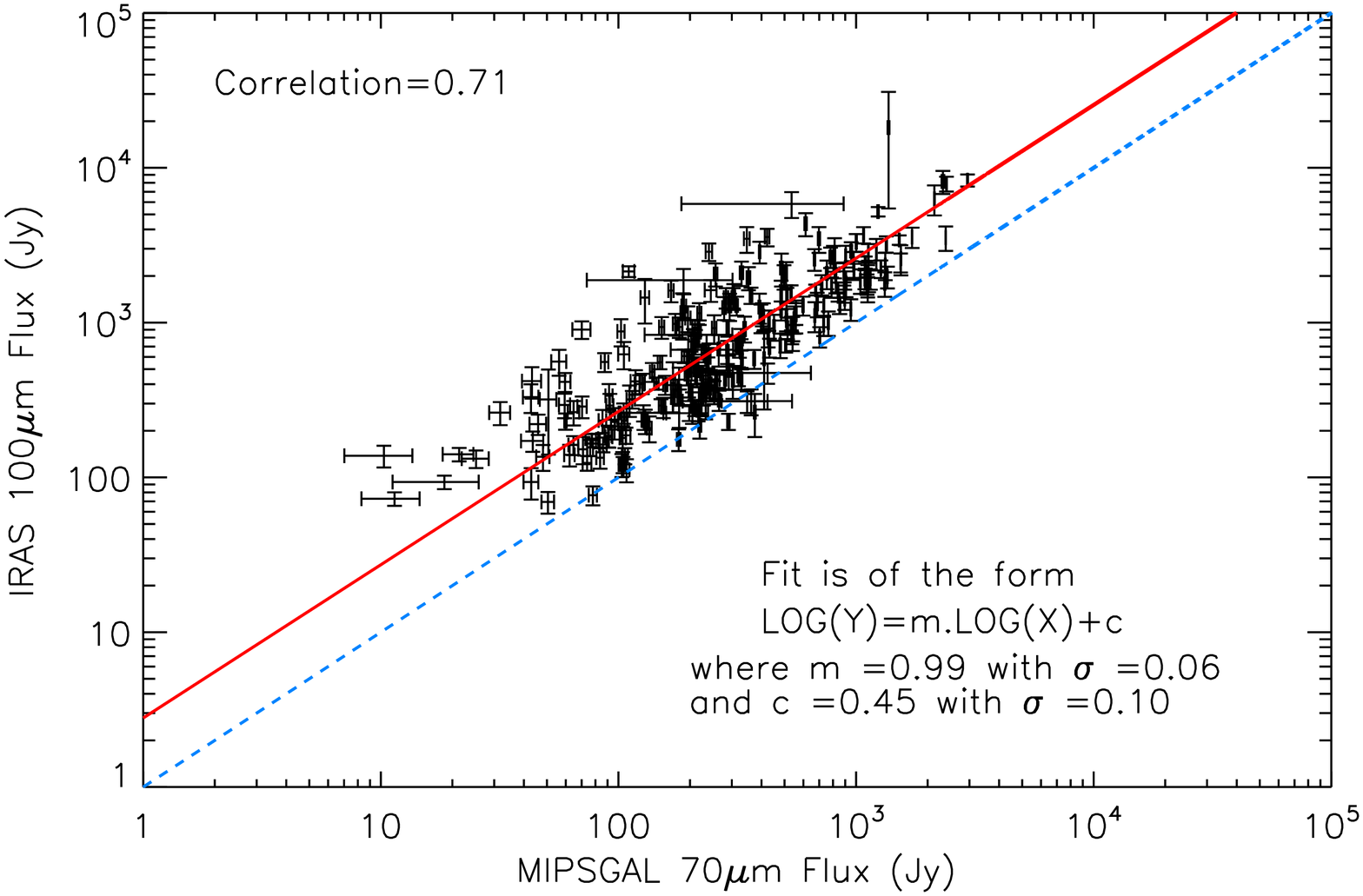}
\caption{A comparison of measured MIPSGAL 70~\micron{} fluxes with IRAS 60~\micron{} (left) and 100~\micron{} (right) PSC fluxes for 243 sources with IRAS PSC detections within 0.7\arcmin{} of the MSX target. The solid line shows the linear-bisector least-squares fit to the data while the dashed line indicates the line of concordance.}
\label{F:discussion_irasvmips_phot}
\end{figure*}

Figure~\ref{F:discussion_irasvmips_phot} shows the comparison between the measured MIPSGAL 70~\micron{} photometry and IRAS PSC 60~\micron{} (left) and 100~\micron{} (right) fluxes for all 242 sources with measurable MIPSGAL fluxes and detections at both 60~\micron{} and 100~\micron{} in the IRAS PSC within 0.7\arcmin{}. Sources which have upper limit IRAS PSC fluxes (i.e. $\sigma$~=~100$\%$) at either 60~\micron{} or 100~\micron{} are not included. Linear bisector least-squares fits to the data give relationships between the MIPSGAL 70~\micron{} and IRAS fluxes of the form shown in equation~\ref{E:data_mipsgal_nonlinearity1}, with powers equal to one within the errors and flux ratios of F$_{IRAS~60}$~$/$~F$_{MIPS~70}$~$\approx$~1.32$\pm$0.21 and F$_{IRAS~100}$~$/$~F$_{MIPS~70}$~$\approx$~2.82$\pm$0.65. The expected flux ratios assuming a blackbody of temperature 50K (or 40K) are F$_{IRAS~60}$~$/$~F$_{MIPS~70}$~=~1.12 (or 0.95) and F$_{IRAS~100}$~$/$~F$_{MIPS~70}$~=~1.88 (or 2.43), so the observed ratios may be in part due to the differences in filter profiles between IRAS and MIPS. However,  the mean ratios derived from fits to the SED models of RMS sources using the model fitter of \citet[][]{Robitaille2007a} are 0.93 and 1.07 for F$_{IRAS~60}$~$/$~F$_{MIPS~70}$ and F$_{IRAS~100}$~$/$~F$_{MIPS~70}$ respectively (Mottram \etal{}, in prep.). The model fitter convolves the models with the observed filters using the standard colour correction, so the model fluxes and flux ratios are directly comparable to the observations. Though the IRAS filters are much wider than the MIPS filter, this effect is included in the ratios produced by the model fitter as well, so the large ratio between IRAS PSC and MIPSGAL fluxes is probably due to contamination in the larger IRAS beam, particularly in the 100~\micron{} band. The smaller model ratio of F$_{IRAS~60}$~$/$~F$_{IRAS~100}$ than would be expected for a blackbody of relevant temperature is probably the result of the distribution of dust temperatures within the SED models. 

The general spread of sources in figure~\ref{F:discussion_irasvmips_phot} is due to several factors. The SED of all young sources will be similar, but not identical, leading to some fluctuation of the flux ratios. As mentioned above, sources in more crowded regions are likely to have more confused IRAS fluxes, leading to both larger flux ratios and larger variation. For sources closer to the galactic centre, the line of sight of observations passes through more of the galactic plane, so there is more chance of confusion.

\subsection{Comparison of MSX and MIPSGAL colours}
\label{S:discussion_mips_colours}

\begin{figure}
\centering
\includegraphics[width=0.49\textwidth]{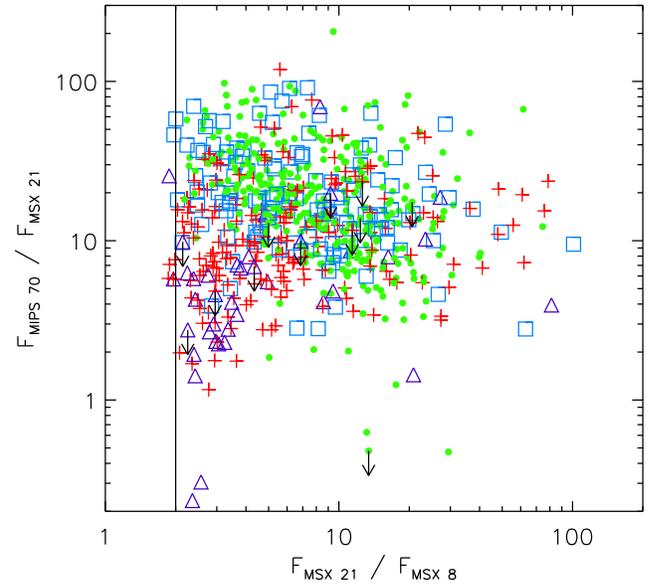}
\caption{Mid and far-IR colour-colour plot using MSX and MIPSGAL 70~\micron{} fluxes. Sources with RMS identifications of `YSO' are shown with `+' signs, `\HII{}~$/$~YSO' with squares, `\HII{} region' with filled circles and `Young~$/$~Old star' with triangles. The vertical solid line indicates the RMS F$_{MSX~21}$~$/$~F$_{MSX~8}$ selection cut, while downward arrows indicate sources with non-detections at 70~\micron{}. Note that sources with fluxes above this cut within the 1~$\sigma$ errors are still included in the sample.}
\label{F:discussion_mips_colours}
\end{figure}

\begin{figure*}
\centering
\includegraphics[width=0.49\textwidth]{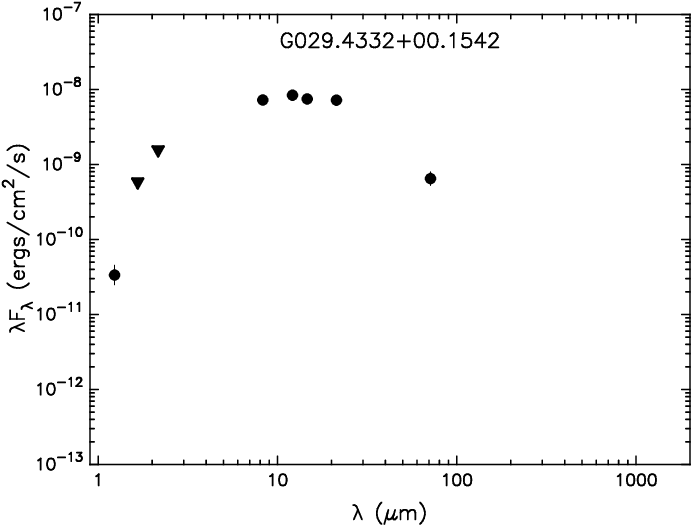}
\includegraphics[width=0.49\textwidth]{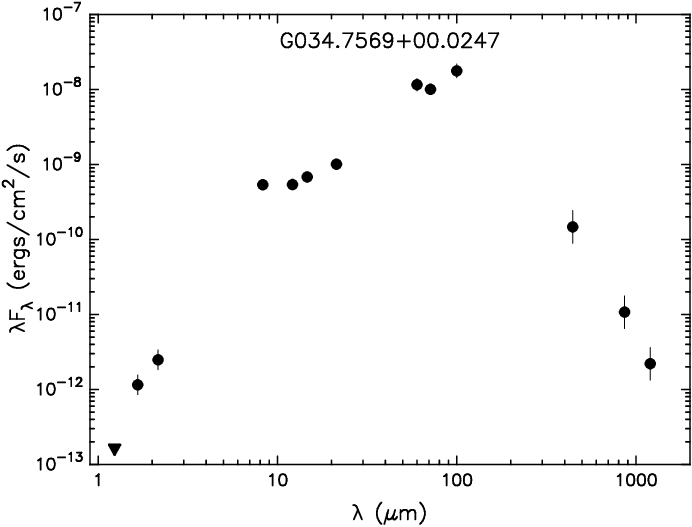}
\caption{The SEDs of G029.4332+00.1542, an evolved star with similar mid-IR colours to young sources and the 'well-behaved' YSO G034.7569+00.0247. The filled circles indicate detected fluxes while the triangles indicate upper limits. The data used include 2MASS, MSX, IRAS, SCUBA sub-millimetre and MAMBO 1.2mm fluxes \citep{Skrutskie2006,Egan2003b,Beichman1988,DiFrancesco2008,Beuther2002a} in addition to those presented in this paper. These and the SEDs of all young RMS sources will be discussed further in Mottram \etal{} in prep.}
\label{F:discussion_mips_seds}
\end{figure*}

Figure~\ref{F:discussion_mips_colours} shows a colour-colour plot using MIPSGAL and MSX fluxes for all 685 sources which have MIPS 70~\micron{} detections and F$_{MSX~8}$~$>$~0 in terms of RMS identification (see \S\ref{S:data_rms}). While young sources have both cold and warm dust associated with them, more evolved sources such as AGB stars will only be associated with strong emission from warm dust \citep[see e.g.][]{vanderVeen1988}. The five sources with F$_{MIPS~70}$~$/$~F$_{MSX~21}$~$<$~1 are therefore unlikely to be MYSOs. Of these, two (G029.4332+00.1542 and G031.9844$-$00.4849) are bright but relatively isolated in GLIMPSE images and so are most likely evolved stars. For comparison the SEDs of G029.4332+00.1542 and G034.7569+00.0247, a typical YSO, are shown in figure~\ref{F:discussion_mips_seds}. It can be seen that inclusion of far-IR flux data is important for discriminating between young and evolved sources. A further two (G034.8625$-$00.0629 and G334.7202+00.1762) appear similar to the previous sources in GLIMPSE images but are also associated with unresolved radio detections, so are most likely PNe. All four sources show $^{13}$CO detections, probably due to diffuse clouds either behind or in front of them. The last (G316.7754$-$00.0447) is certainly in a star forming region as both MSX and GLIMPSE images show multiple sources and diffuse emission nearby. However no point sources are evident in any of the Spitzer IRAC bands, so this source is either a more evolved \HII{} region or a photo-dissociation region (PDR) which has bright mid-IR emission. RMS sources already identified as evolved are not included in this paper, and have not had their MIPSGAL 70~\micron{} fluxes measured.

In general the \HII{} regions have larger F$_{MSX~21}$~$/$~F$_{MSX~8}$ and F$_{MIPS~70}$~$/$~F$_{MSX~21}$ ratios than the YSOs, which is consistent with the \HII{} regions being more embedded than the YSOs. However, there is a population of YSOs which have similar colours to the \HII{} regions.

\section{Summary and conclusions}
\label{S:conclusions}

In this paper, we have presented a new technique for background subtraction in aperture photometry using a 2-D fit to the background within the sky annulus. This was then used to obtain photometry from MIPSGAL 70~\micron{} mosaic images for 724 young RMS sources and for 734 such sources using IRAS Galaxy Atlas 60~\micron{} and 100~\micron{} images. Overall, far-infrared fluxes have been obtained for 1113 of the 1336 candidates examined, of which 370 have no IRAS PSC entry and a further 374 have only upper limits in the IRAS PSC at 60~\micron{} and$/$or 100~\micron{}. In addition, 4 objects have been identified as evolved due to their MIPS fluxes. This is one of the first uses of MIPSGAL 70~\micron{} data (another example being the work of \citet{Chapin2008}) and one of the largest sets of measurements of far-IR sources in the galactic plane. Comparisons with the IRAS PSC, where applicable, show reasonable agreement, though the MIPSGAL fluxes are strictly superior due to improved resolution, and the IGA data includes a hysteresis correction at 60~\micron{} which the IRAS PSC did not.

These far-IR flux measurements, along with other photometric data, provide information about the spectral energy distribution (SED) of young RMS sources, and thus allow the determination of the luminosities of this sample of sources. This will allow the systematic study of the properties of a large, well-selected sample of MYSOs as a function of luminosity, and will be the subject of forthcoming papers. Information on all sources discussed in this paper is available via the RMS database, which can be found at www.ast.leeds.ac.uk/RMS/.

\begin{acknowledgements}

The authors would like to thank the anonymous referee for comments and suggestions which improved the clarity of this paper. We would like to thank Chad Engelbracht and Roberta Paladini for their helpful discussions regarding pixel non-linearity correction of MIPSGAL images. We also thank Davy Kirkpatrick at IPAC for his help obtaining all IGA images and beam maps. JCM is partially funded by a Postgraduate Studentship and by a Postdoctoral fellowship from the Science and Technologies Research Council of the United Kingdom (STFC).

\end{acknowledgements}

\bibliographystyle{./aa}

\begin{thebibliography}{36}
\expandafter\ifx\csname natexlab\endcsname\relax\def\natexlab#1{#1}\fi

\bibitem[{{Baker} \& {Menzel}(1938)}]{Baker1938}
{Baker}, J.~G. \& {Menzel}, D.~H. 1938, \apj, 88, 52

\bibitem[{{Beichman} {et~al.}(1988){Beichman}, {Neugebauer}, {Habing}, {Clegg},
  \& {Chester}}]{Beichman1988}
{Beichman}, C.~A., {Neugebauer}, G., {Habing}, H.~J., {Clegg}, P.~E., \&
  {Chester}, T.~J., eds. 1988, {Infrared astronomical satellite (IRAS) catalogs
  and atlases}, Vol. 1: Explanatory supplement

\bibitem[{{Benjamin} {et~al.}(2003){Benjamin}, {Churchwell}, {Babler}, {Bania},
  {Clemens}, {Cohen}, {Dickey}, {Indebetouw}, {Jackson}, {Kobulnicky},
  {Lazarian}, {Marston}, {Mathis}, {Meade}, {Seager}, {Stolovy}, {Watson},
  {Whitney}, {Wolff}, \& {Wolfire}}]{Benjamin2003}
{Benjamin}, R.~A., {Churchwell}, E., {Babler}, B.~L., {et~al.} 2003, \pasp,
  115, 953

\bibitem[{{Berriman} {et~al.}(2006){Berriman}, {Laity}, {Good}, {Jacob},
  {Katz}, {Deelman}, {Singh}, {Su}, \& {Prince}}]{Berriman2006}
{Berriman}, G.~B., {Laity}, A.~C., {Good}, J.~C., {et~al.} 2006, in The Far
  Infrared and Submillimetre Universe, Proceedings of Earth Sciences Technology
  Conference

\bibitem[{{Beuther} {et~al.}(2002{\natexlab{a}}){Beuther}, {Schilke}, {Gueth},
  {McCaughrean}, {Andersen}, {Sridharan}, \& {Menten}}]{Beuther2002c}
{Beuther}, H., {Schilke}, P., {Gueth}, F., {et~al.} 2002{\natexlab{a}}, \aap,
  387, 931

\bibitem[{{Beuther} {et~al.}(2002{\natexlab{b}}){Beuther}, {Schilke}, {Menten},
  {Motte}, {Sridharan}, \& {Wyrowski}}]{Beuther2002a}
{Beuther}, H., {Schilke}, P., {Menten}, K.~M., {et~al.} 2002{\natexlab{b}},
  \apj, 566, 945

\bibitem[{{Beuther} {et~al.}(2002{\natexlab{c}}){Beuther}, {Schilke},
  {Sridharan}, {Menten}, {Walmsley}, \& {Wyrowski}}]{Beuther2002b}
{Beuther}, H., {Schilke}, P., {Sridharan}, T.~K., {et~al.} 2002{\natexlab{c}},
  \aap, 383, 892

\bibitem[{{Bevington}(1969)}]{Bevington1969}
{Bevington}, P.~R. 1969, {Data reduction and error analysis for the physical
  sciences}, 1st edn. (McGraw-Hill)

\bibitem[{{Cao} {et~al.}(1997){Cao}, {Terebey}, {Prince}, \&
  {Beichman}}]{Cao1997}
{Cao}, Y., {Terebey}, S., {Prince}, T.~A., \& {Beichman}, C.~A. 1997, \apjs,
  111, 387

\bibitem[{{Carey} {et~al.}(2009){Carey}, {Noriega-Crespo}, {Mizuno}, {Shenoy},
  {Paladini}, {Kraemer}, {Price}, {Flagey}, {Ryan}, {Ingalls}, {Kuchar},
  {Pinheiro Gon{\c c}alves}, {Indebetouw}, {Billot}, {Marleau}, {Padgett},
  {Rebull}, {Bressert}, {Ali}, {Molinari}, {Martin}, {Berriman}, {Boulanger},
  {Latter}, {Miville-Deschenes}, {Shipman}, \& {Testi}}]{Carey2009}
{Carey}, S.~J., {Noriega-Crespo}, A., {Mizuno}, D.~R., {et~al.} 2009, \pasp,
  121, 76

\bibitem[{{Chapin} {et~al.}(2008){Chapin}, {Ade}, {Bock}, {Brunt}, {Devlin},
  {Dicker}, {Griffin}, {Gundersen}, {Halpern}, {Hargrave}, {Hughes}, {Klein},
  {Marsden}, {Martin}, {Mauskopf}, {Netterfield}, {Olmi}, {Pascale},
  {Patanchon}, {Rex}, {Scott}, {Semisch}, {Truch}, {Tucker}, {Tucker}, {Viero},
  \& {Wiebe}}]{Chapin2008}
{Chapin}, E.~L., {Ade}, P.~A.~R., {Bock}, J.~J., {et~al.} 2008, \apj, 681, 428

\bibitem[{{Churchwell} {et~al.}(2009){Churchwell}, {Babler}, {Meade},
  {Whitney}, {Benjamin}, {Indebetouw}, {Cyganowski}, {Robitaille}, {Povich},
  {Watson}, \& {Bracker}}]{Churchwell2009}
{Churchwell}, E., {Babler}, B.~L., {Meade}, M.~R., {et~al.} 2009, \pasp, 121,
  213

\bibitem[{{Clarke} {et~al.}(2006){Clarke}, {Lumsden}, {Oudmaijer}, {Busfield},
  {Hoare}, {Moore}, {Sheret}, \& {Urquhart}}]{Clarke2006}
{Clarke}, A.~J., {Lumsden}, S.~L., {Oudmaijer}, R.~D., {et~al.} 2006, \aap,
  457, 183

\bibitem[{{Dale} {et~al.}(2007){Dale}, {Gil de Paz}, {Gordon}, {Hanson},
  {Armus}, {Bendo}, {Bianchi}, {Block}, {Boissier}, {Boselli}, {Buckalew},
  {Buat}, {Burgarella}, {Calzetti}, {Cannon}, {Engelbracht}, {Helou},
  {Hollenbach}, {Jarrett}, {Kennicutt}, {Leitherer}, {Li}, {Madore}, {Martin},
  {Meyer}, {Murphy}, {Regan}, {Roussel}, {Smith}, {Sosey}, {Thilker}, \&
  {Walter}}]{Dale2007}
{Dale}, D.~A., {Gil de Paz}, A., {Gordon}, K.~D., {et~al.} 2007, \apj, 655, 863

\bibitem[{{Di Francesco} {et~al.}(2008){Di Francesco}, {Johnstone}, {Kirk},
  {MacKenzie}, \& {Ledwosinska}}]{DiFrancesco2008}
{Di Francesco}, J., {Johnstone}, D., {Kirk}, H., {MacKenzie}, T., \&
  {Ledwosinska}, E. 2008, \apjs, 175, 277

\bibitem[{{Egan} {et~al.}(2003){Egan}, {Price}, \& {Kraemer}}]{Egan2003b}
{Egan}, M.~P., {Price}, S.~D., \& {Kraemer}, K.~E. 2003, \baas, 35, 1301

\bibitem[{{Egan} {et~al.}(1999){Egan}, {Price}, {Moshir}, {Cohen}, \&
  {Tedesco}}]{Egan1999}
{Egan}, M.~P., {Price}, S.~D., {Moshir}, M.~M., {Cohen}, M., \& {Tedesco}, E.
  1999, NASA STI/Recon Technical Report, 14854

\bibitem[{{Engelbracht} {et~al.}(2007){Engelbracht}, {Blaylock}, {Su}, {Rho},
  {Rieke}, {Muzerolle}, {Padgett}, {Hines}, {Gordon}, {Fadda},
  {Noriega-Crespo}, {Kelly}, {Latter}, {Hinz}, {Misselt}, {Morrison},
  {Stansberry}, {Shupe}, {Stolovy}, {Wheaton}, {Young}, {Neugebauer},
  {Wachter}, {P{\'e}rez-Gonz{\'a}lez}, {Frayer}, \&
  {Marleau}}]{Engelbracht2007}
{Engelbracht}, C.~W., {Blaylock}, M., {Su}, K.~Y.~L., {et~al.} 2007, \pasp,
  119, 994

\bibitem[{{Gordon} {et~al.}(2007){Gordon}, {Engelbracht}, {Fadda},
  {Stansberry}, {Wachter}, {Frayer}, {Rieke}, {Noriega-Crespo}, {Latter},
  {Young}, {Neugebauer}, {Balog}, {Beeman}, {Dole}, {Egami}, {Haller}, {Hines},
  {Kelly}, {Marleau}, {Misselt}, {Morrison}, {P{\'e}rez-Gonz{\'a}lez}, {Rho},
  \& {Wheaton}}]{Gordon2007}
{Gordon}, K.~D., {Engelbracht}, C.~W., {Fadda}, D., {et~al.} 2007, \pasp, 119,
  1019

\bibitem[{{Henning} {et~al.}(1984){Henning}, {Friedemann}, {G{\"u}rtler}, \&
  {Dorschner}}]{Henning1984}
{Henning}, T., {Friedemann}, C., {G{\"u}rtler}, J., \& {Dorschner}, J. 1984,
  \an, 305, 67

\bibitem[{{Hoare} {et~al.}(2005){Hoare}, {Lumsden}, {Oudmaijer}, {Urquhart},
  {Busfield}, {Sheret}, {Clarke}, {Moore}, {Allsopp}, {Burton}, {Purcell},
  {Jiang}, \& {Wang}}]{Hoare2005}
{Hoare}, M.~G., {Lumsden}, S.~L., {Oudmaijer}, R.~D., {et~al.} 2005, in IAU
  Symposium, Vol. 227, Massive Star Birth: A Crossroads of Astrophysics, ed.
  R.~{Cesaroni}, M.~{Felli}, E.~{Churchwell}, \& M.~{Walmsley}, 370--375

\bibitem[{{Lumsden} {et~al.}(2002){Lumsden}, {Hoare}, {Oudmaijer}, \&
  {Richards}}]{Lumsden2002}
{Lumsden}, S.~L., {Hoare}, M.~G., {Oudmaijer}, R.~D., \& {Richards}, D. 2002,
  \mnras, 336, 621

\bibitem[{{Miville-Desch{\^e}nes} \& {Lagache}(2005)}]{MivilleDeschenes2005}
{Miville-Desch{\^e}nes}, M.-A. \& {Lagache}, G. 2005, \apjs, 157, 302

\bibitem[{{Mottram}(2008)}]{Mottram2008}
{Mottram}, J.~C. 2008, PhD thesis, University of Leeds, UK

\bibitem[{{Mottram} {et~al.}(2007){Mottram}, {Hoare}, {Lumsden}, {Oudmaijer},
  {Urquhart}, {Sheret}, {Clarke}, \& {Allsopp}}]{Mottram2007}
{Mottram}, J.~C., {Hoare}, M.~G., {Lumsden}, S.~L., {et~al.} 2007, \aap, 476,
  1019

\bibitem[{{Mottram} {et~al.}(2006){Mottram}, {Urquhart}, {Hoare}, {Lumsden}, \&
  {Oudmaijer}}]{Mottram2006}
{Mottram}, J.~C., {Urquhart}, J.~S., {Hoare}, M.~G., {Lumsden}, S.~L., \&
  {Oudmaijer}, R.~D. 2006, ArXiv:astro-ph/0612481

\bibitem[{{Robitaille} {et~al.}(2007){Robitaille}, {Whitney}, {Indebetouw}, \&
  {Wood}}]{Robitaille2007a}
{Robitaille}, T.~P., {Whitney}, B.~A., {Indebetouw}, R., \& {Wood}, K. 2007,
  \apjs, 169, 328

\bibitem[{{Skrutskie} {et~al.}(2006){Skrutskie}, {Cutri}, {Stiening},
  {Weinberg}, {Schneider}, {Carpenter}, {Beichman}, {Capps}, {Chester},
  {Elias}, {Huchra}, {Liebert}, {Lonsdale}, {Monet}, {Price}, {Seitzer},
  {Jarrett}, {Kirkpatrick}, {Gizis}, {Howard}, {Evans}, {Fowler}, {Fullmer},
  {Hurt}, {Light}, {Kopan}, {Marsh}, {McCallon}, {Tam}, {Van Dyk}, \&
  {Wheelock}}]{Skrutskie2006}
{Skrutskie}, M.~F., {Cutri}, R.~M., {Stiening}, R., {et~al.} 2006, \aj, 131,
  1163

\bibitem[{{Urquhart} {et~al.}(2007{\natexlab{a}}){Urquhart}, {Busfield},
  {Hoare}, {Lumsden}, {Clarke}, {Moore}, {Mottram}, \&
  {Oudmaijer}}]{Urquhart2007a}
{Urquhart}, J.~S., {Busfield}, A.~L., {Hoare}, M.~G., {et~al.}
  2007{\natexlab{a}}, \aap, 461, 11

\bibitem[{{Urquhart} {et~al.}(2007{\natexlab{b}}){Urquhart}, {Busfield},
  {Hoare}, {Lumsden}, {Oudmaijer}, {Moore}, {Gibb}, {Purcell}, {Burton}, \&
  {Marechal}}]{Urquhart2007c}
{Urquhart}, J.~S., {Busfield}, A.~L., {Hoare}, M.~G., {et~al.}
  2007{\natexlab{b}}, \aap, 474, 891

\bibitem[{{Urquhart} {et~al.}(2008{\natexlab{a}}){Urquhart}, {Busfield},
  {Hoare}, {Lumsden}, {Oudmaijer}, {Moore}, {Gibb}, {Purcell}, {Burton},
  {Mar{\'e}chal}, {Jiang}, \& {Wang}}]{Urquhart2008a}
{Urquhart}, J.~S., {Busfield}, A.~L., {Hoare}, M.~G., {et~al.}
  2008{\natexlab{a}}, \aap, 487, 253

\bibitem[{{Urquhart} {et~al.}(2008{\natexlab{b}}){Urquhart}, {Hoare},
  {Lumsden}, {Oudmaijer}, \& {Moore}}]{Urquhart2008b}
{Urquhart}, J.~S., {Hoare}, M.~G., {Lumsden}, S.~L., {Oudmaijer}, R.~D., \&
  {Moore}, T.~J.~T. 2008{\natexlab{b}}, in Astronomical Society of the Pacific
  Conference Series, Vol. 387, Massive Star Formation: Observations Confront
  Theory, ed. H.~{Beuther}, H.~{Linz}, \& T.~{Henning}, 381

\bibitem[{{Urquhart} {et~al.}(2009){Urquhart}, {Hoare}, {Purcell}, {Lumsden},
  {Oudmaijer}, {Moore}, {Busfield}, {Mottram}, \& {Davies}}]{Urquhart2009}
{Urquhart}, J.~S., {Hoare}, M.~G., {Purcell}, C.~R., {et~al.} 2009, \aap, 501,
  539

\bibitem[{{van der Veen} \& {Habing}(1988)}]{vanderVeen1988}
{van der Veen}, W.~E.~C.~J. \& {Habing}, H.~J. 1988, \aap, 194, 125

\bibitem[{{Wheelock} {et~al.}(1994){Wheelock}, {Gautier}, {Chillemi}, {Kester},
  {McCallon}, {Oken}, {White}, {Gregorich}, {Boulanger}, \&
  {Good}}]{Wheelock1994}
{Wheelock}, S.~L., {Gautier}, T.~N., {Chillemi}, J., {et~al.} 1994, NASA
  STI/Recon Technical Report, 95, 22539

\bibitem[{{Wynn-Williams}(1982)}]{Wynn-Williams1982}
{Wynn-Williams}, C.~G. 1982, \araa, 20, 587

\end{thebibliography}

\label{lastpage}

\end{document}